# Cashing Out: Assessing the risk of localised financial exclusion as the UK moves towards a cashless society

**George Sullivan & Luke Burns**

University of Leeds, Leeds, United Kingdom

## Abstract

Whilst academic, commercial and policy literature on financial exclusion is extensive and wide-ranging, there have been very few attempts to quantify and measure localised financial exclusion anywhere in the world. This is a subject of growing importance in modern UK society with the withdrawal of cash infrastructure and a shift towards online banking. This research develops a composite indicator using a wide-range of input variables, including the locations of existing cash infrastructure, various demographic factors (such as income and housing tenure) and other freely available lifestyle data to identify areas at greatest risk of financial exclusion, thereby aiding organisations to develop intervention strategies to tackle the problem. The indicator illustrates that whilst there is no apparent correlation between financial exclusion and deprivation, pockets of extreme financial exclusion are generally found in deprived communities, and affluent, suburban areas tend to score consistently more favourably and consequently carry less risk. The attributing causes vary, from a lack of infrastructure, to low car availability, but income levels have a pronounced influence. Three policy proposals are put forward, including offering banking services at PayPoint outlets, and converting cash machines to cash recyclers, but improving digital adoption was found to be the most effective intervention, provided that it is implemented by community organisations. Policies purely targeting infrastructure provision or addressing social exclusion are unlikely to be effective, as community-based initiatives coupled with wider reforms to the financial system are needed.

## Key Words

Financial Exclusion, Cashless Society, Banking, Financial Services, Composite Index

# Background and Rationale

With the UK moving towards a cashless society, certain communities are at risk of being left behind, not only from the financial system but also wider society. Cash provision and online banking usage are heavily influenced by localised demographic factors and behaviour, thus this research will [1] propose a framework whereby areas most at risk can be identified, [2] understand why certain areas are more affected than others and [3] suggest tailored intervention strategies to avoid full-scale financial and societal exclusion.

The UK could be a cashless society by 2026, presenting a major challenge for local and national government given that 2.2 million people are reliant on cash, and 1.3 million do not hold an active bank account (Ceeney, 2019). Furthermore, the UK is seeing a national trend towards Automatic Teller Machines (ATMs) closing or being converted to fee-charging machines, whilst stronger identification and security rules to tackle money laundering make bank accounts more difficult to open. This changing landscape relates to wider political and academic discussion of accelerating the rates of bank branch closures, the growth of digital payments and the role of the Post Office (PO) in providing financial services.

Academic literature on cash access has largely focused on the role of bank branches from an economic geography perspective, but other than New Labour's social exclusion agenda in the late 1990's, there has been little work on the wider impact of financial exclusion. Recent work from the University of Bristol's Personal Finance Research Centre (PFRC) has attempted to study the effect of banking without branches and quantify cash access through an innovative 'Av Cash' index (Tischer et al., 2019).

This research will build on the work of the PFRC by developing a risk index, based on infrastructure provision, the demographics of cash use and financial provision and the availability of alternatives. This model will allow communities, banks and policymakers to identify the area's most at risk of financial exclusion, understand the implications of this and design effective policies to ensure citizens aren't left behind. The index set out in this paper is designed to be easily replicable and transferable to different regions within the UK. In the context of this research, the index is evidenced on the city of Nottingham, in the Midlands, United Kingdom (UK).

# Financial Exclusion in the UK

Financial exclusion research was spurred by New Labour's social exclusion agenda in the 1990's, much of it completed by Kempson et al. (2000), and whilst it has been criticised (notably by French et al. (2008)) for being produced largely for the benefit of the British Bankers Association and not adequately considering geographical dimensions, it is a useful starting point for understanding financial exclusion today. Financial exclusion is generally defined as *"the inability, difficulty or reluctance of particular groups to access mainstream financial services"* (McKillopand Wilson, 2007, p.9).

Historical change since the 1980's has caused financial exclusion to become more pronounced in parts of the population (Kempson et al., 2000). Firstly, rising wage inequalities (due to higher skills and a decline in manufacturing) has resulted in fewer wages, pensions and benefits paid in cash, with a greater reliance on bank accounts. This is coupled with a more flexible labour market that doesn't always guarantee a fixed income, making it difficult to apply for financial products. Secondly, demographic change has resulted in more single parents, increasing the level of financial dependence, whilst as the population ages, a growing gap has emerged between pensioners with extensive assets and those reliant on cash and state pensions. Finally, the housing market reform has resulted in increased home ownership (Devlin, 2005), whilst there has been a growing concentration of people on low incomes in social and private rental housing. Such an environment of financial exclusion can also result in a self-reinforcing cycle. Areas of social housing can result in higher insurance premiums, whilst the lack of a bank account reduces access to other products and services (Hogarth and O'Donnell, 2000).

Statistical analysis undertaken by Devlin (2005) on the likelihood of holding financial products by socio- economic circumstances broadly mirrored the findings of Kempson et al. (2000), in particular with reference to their qualitative research on bank account access. Whilst Kempson et al. (2000) found those without bank accounts concentrated amongst those on low incomes, the unemployed, lone parents and retirees (see Table 1), Devlin (2005) found the key influences of financial exclusion to be employment status, household income and housing tenure. As Table 1 suggests, low income can also make savings and insurance products unaffordable, further decreasing participation in the system. Kempson et al. (2000) also note the importance of making a distinction between those disengaged from the banking system (due to being recently out of work, retired, or living with a long- term health condition), and those who are denied access to an account. These are likely to be living off savings, relying solely on their partner's account, face cultural and religious barriers to opening an account, or have been denied by banks (due to a lack of ID or failed background or credit checks) – this is particularly common amongst the homeless or new migrants to the UK.

Table 1 presents a thorough review of the factors deemed to influence financial exclusion, as retrieved following an in-depth and academic and policy literature search.

| Cause | Commentary (with relevant literature bolded for emphasis) |
|---|---|
| **Age** (Slight influence) | Those aged 66+ significantly less likely to have a current account than others **(Devlin, 2005)**, whilst **Kempson et al. (2000)** is equivocal on the role of age. **The Select Committee on Financial Exclusion (2017)** also saw increased financial exclusion amongst young people, though other than the **Hogarth and O'Donnell (2000)** US study, there is little empirical evidence to support this. |
| **Digital Literacy** (Growing influence) | Whilst this wasn't relevant (or timely) when the majority of financial exclusion research was undertaken, it is a notable growing influence today, in a market facing rising branch closures and a shift in attitudes to online banking **(Ceeney, 2019; Ripley and Watmough, 2020; Select Committee on Financial Exclusion, 2017)**. |
| **Educational attainment** (Slight influence) | Those with no formal qualifications (or only up to GCSE level) are likely to be excluded, due to low knowledge of financial services **(Devlin, 2005)**, although its influence was highly correlated with those out of work and on low incomes. |
| **Employment status** (Pronounced influence) | Found by many to be a pronounced and consistent influence, largely as a bank account is needed to receive wages, whilst a regular income is required to apply for many financial products **(Devlin, 2005; Devlin, 2009; Hogarth and O'Donnell, 2000; Kempson et al., 2000; Select Committee on Financial Exclusion, 2017)**. |
| **Ethnicity** (Secondary influence) | Whilst there is little evidence that ethnicity is an influence on financial exclusion, ethnic minority groups are more prevalent amongst those on low incomes **(Kempson et al., 2000)**. |
| **Geography** (Pronounced influence) | Some surveys have found variation across regions **(Kempson et al., 2000)**, whilst others have found the variation not to influence exclusion **(Devlin, 2009)**. At the local scale, the influence is more visible, with infrastructure provision found to be poorer in lower-income areas **(Leyshon et al., 2008)**. Despite only 1% of those without a current account attributing it to not living near a branch **(Kempson and Whyley, 1999)**, access to cash is inherently spatial **(Tischer et al., 2019)**. |
| **Health** (Secondary influence) | Health may mean there is exclusion from certain financial services, such as branches for those with mobility issues, whilst those with limiting long-term illness may be unable to access work **(Kempson et al., 2000)**, however, the exact influence health has on financial exclusion is difficult to quantify, with the census' measure of health being subjective. |
| **Housing tenure** (Pronounced influence) | Multiple studies found housing tenure to be a consistent and marked influence on financial exclusion **(Devlin, 2005; Devlin, 2009; Kempson and Whyley, 1999)**. Whilst it is generally correlated with income, housing tenure, in particular, was found to develop pockets of financial exclusion on marginalised council estates **(Kempson and Whyley, 1999)**. |
| **Income** (Most pronounced) | Possibly the most pronounced, with financial exclusion having a strong association with poverty, as well as reducing the ability to hold other financial products **(Devlin, 2005)**. **Kempson et al. (2000)** found that only 35% of those on income support (now part of Universal Credit) owned a financial product. Income was also the most pronounced factor in studies by **Hogarth and O'Donnell (2000) and Caskey (1997)**. |
| **Lone parenthood** (Some influence) | **Kempson and Whyley (1999)** found lone-parent households are very low users of financial services, with a quarter having no products, and tend to be concentrated on low incomes **(Devlin, 2005)**. Whilst this factor is not always in correlation with the more pronounced variables, there is a greater risk of deeper financial exclusion within this group **(Kempson et al., 2000)**. |

*Table 1: Summary of causes of financial exclusion.*

# Cash Use and Operations in the UK

According to Ceeney (2019), cash use in the UK has halved in the past ten years and is expected to halve again over the upcoming ten years. As shown in Figure 1, there are various methods of accessing cash available to the general public, but ATMs remain the preferred method, accounting for 97% of all withdrawals (Ceeney, 2019), even though increasing numbers are being converted to fee-charging machines (Which?, 2019). Once the mainstay of provision, bank branches have been closing since the 1980s (French et al., 2008) and this is the preferred cash access method for just 7% of consumers (Britain Thinks, 2019). Whilst Post Offices and supermarkets could preserve cash provision, it's only the preferred option for 5% and 4% of consumers respectively, and Parrott (2018) found that only 2 in 5 people (40%) are aware of the Post Office's financial offering.

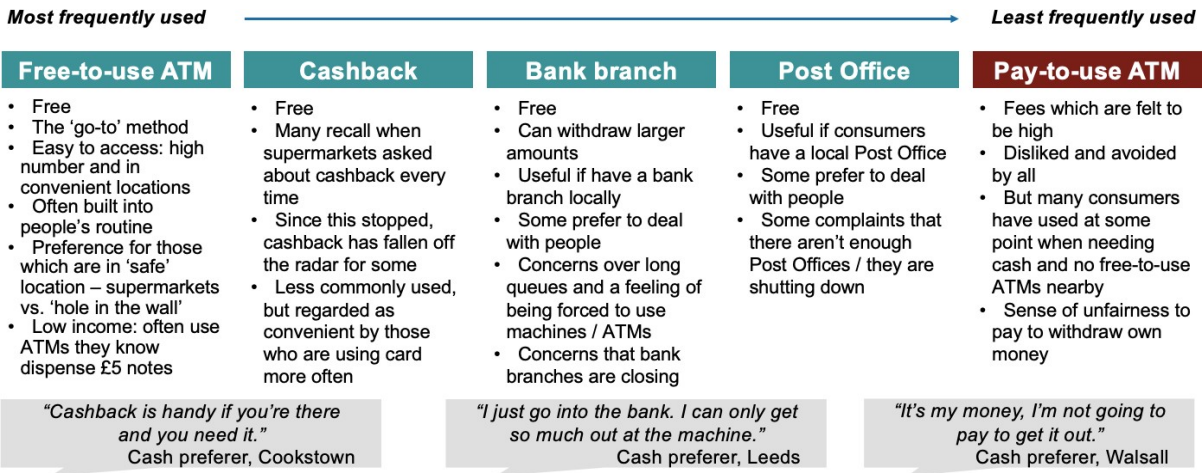

*Figure 1: Analysis from Britain Thinks (2019) on the methods of accessing cash.*

Toynbee Hall (2013) and Britain Thinks (2019) found those from a low socio-economic background, with a long-term health condition, and those who are digitally excluded are more likely to prefer cash. Whilst older generations are more likely to use cash, it's not straightforward, as digital access is a greater determinant of use. Moreover, motivations to use cash are driven by its ability to help budget, to help consumers remain in financial control and to maintain entrenched habits. The preference for cash use depends on a person's openness to the digital world. Cash preferers tend to have security concerns over digital payments, fear of unknown technology, issues with the transparency of card payments, and a deep distrust of the banking system. Consequently, heavy cash users will feel little motivation to switch to digital payments (Britain Thinks, 2019).

Despite living in challenging times with regards to COVID-19, Auer et al. (2020) show infection transmission from banknotes is low when compared to other payment methods, yet perceptions have seen businesses stop accepting cash and the UK contactless limit raised to £100, partly in response to seeing the value of ATM withdrawals falling (UK Finance, 2021). Whilst the Bank of England (2020) attempted to bolster confidence in banknotes, behaviour changes could promote a more rapid decline of cash, potentially opening up a payment divide as it becomes increasingly difficult to access and use.

Figure 2 illustrates the UK's complex and fragmented cash infrastructure model. Ceeney (2019) identified the printing, wholesale and distribution of cash elements are owned by separate companies, designed around large transaction volumes, costing between £5 and £9 billion per year to operate. As cash use falls, the viability of this model is threatened, and thus a small failure can impact the entire system.

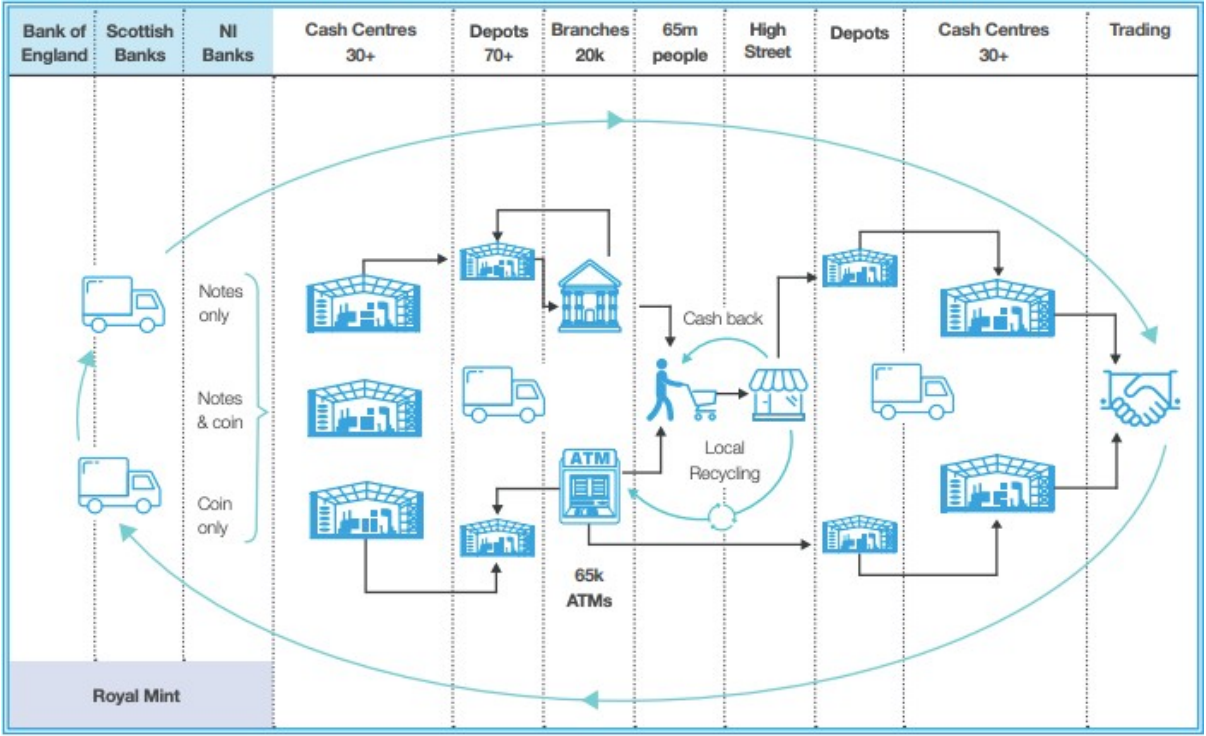

*Figure 2: The cash cycle in the UK, from Ceeney (2019).*

The UK has codes of practice to try and preserve the cash model. When a customer uses another bank's free ATM, their bank pays the operator an interchange fee. LINK (an inter-bank ATM scheme) has a protected ATM programme, whereby certain ATMs in deprived areas (or a fair distance from a nearby ATM) have higher fees, subsidising unprofitable machines. Operators that set up new free-to-use ATMs or convert fee-charging ATMs will benefit from a 'super-premium' of £2.75, rather than the low 25p per withdrawal, in certain areas (LINK, 2019).

In 2015, UK banks signed up to an industry-wide Access to Banking Standard, including giving at least three months' notice before closing a branch, publishing an impact assessment, and contacting vulnerable customers to support them with alternative ways to bank. A review by Griggs (2016) was largely supportive of the banks' positions, viewing them as businesses rather than utility providers. He found that decisions made were largely driven by market pressures, rather than cost-cutting, and the effects of geographic delineation and growth of online were being felt by all businesses. However, the review recommended that banks should work far more closely with older and vulnerable customers in the future.

## Geographical Dimensions of Cash Access

Following the recession in the 1990's, UK banks became more risk-averse and began to withdraw from disadvantaged communities (Leyshon and Thrift, 1994), retreating to a more affluent customer base at the expense of poorer consumers. However, French et al. (2008) found that discussions surrounding branch closures had largely been erased from academic and political discussion. Since then, the financial crisis aftermath and growth in digital banking have seen around 55 branches close each month since 2015, with potentially more at risk as a result of COVID-19-induced falls in demand (Which?, 2020).

Work by Leyshon et al. (2008) in the Welsh valleys found closures disproportionately concentrated within poorer areas, with policy failing to consider the uneven geography of financial services. The key drivers behind these closures were mergers, neoliberal financial reforms, new distribution channels (e.g. telephone banking) and the demutualisation of building societies. The geographies of bank networks pre-rationalisation were skewed towards certain economic geographies. As the economy changed and networks adjusted to it, closures were exacerbated in deprived metropolitan areas.

Statistical analysis carried out by Leyshon et al. (2008) found a broad relationship between closure rates and deprivation (though there was a lower closure rate for building societies). The areas worst affected were identified as 'multicultural metropolitan', 'traditional manufacturing' and 'built-up areas' in the ONS' Area Classification of 'Super Groups'. Later research by French et al. (2020) and Which? (2020) found this trend has continued into the 21st century, but with increased geographic variation.

Since 2003, Post Offices have become a key component of banking provision. By introducing universal banking services, the government aimed to modernise welfare by making payments directly into bank accounts, providing access to everyone who wants a bank account at Post Office counters and finding new customers to keep the network viable (Midgley, 2005). This led to basic bank accounts (with no overdraft or cheque books) being planned to be provided by the Post Office. However, a fear that this would create a "*poor person's bank*" (Midgley, 2005, p.279) saw these accounts provided through high street banks with access over a Post Office counter. Since then, the Post Office has developed its own range of financial services in partnership with the Bank of Ireland, with a 'card account' for receiving benefit payments for those denied a traditional bank account. However, the Department for Work and Pensions (DWP) opted not to renew this contract from 2021, meaning many benefit claimants are now seeking to open bank accounts where possible (Jefferies, 2020).

Universal coverage of the Post Office has been tested somewhat by the Network Change Programme (Figure 3) and the Urban Network Reinvention Programme. Whilst Post Offices have been in decline since the 1960's, the Urban Network Reinvention Programme from 2002 to 2005 rationalised 2,500 branches in urban areas, largely on a straight-line measure. A more noteworthy restructuring came in the Network Change Programme in 2007, setting criteria of access. However, the House of Commons Business and Enterprise Committee (2008), found loopholes including replacing branches with mobile vans, and the Post Office not understanding whether the national criteria were being met in local areas. Langford and Higgs (2010) and Comber et al. (2009) both found that the criteria were not met before the closures in either the Welsh valleys or Leicester. Loss of access was greater in urban areas, particularly as consumers tend not to behave rationally and travel only to their nearest Post Office, possibly combining a visit with work, or where car parking is available, whilst Macintyre et al. (2008) found in Glasgow, Post Offices were more common in deprived areas.

Table I  Access criteria under the Network Change Programme

| | |
|---|---|
| 1 | 99% of the national population should remain within 3 miles, and 90% should remain within 1 mile, of their nearest post office branch |
| 2 | 99% of the total population in 'deprived urban areas' should remain within 1 mile of their nearest post office branch |
| 3 | 95% of the total urban population should remain within 1 mile of their nearest post office branch |
| 4 | 95% of the total rural population should remain within 3 miles of their nearest post office branch |
| 5 | 95% of the population resident within each postcode district boundary should remain within 6 miles of their nearest post office branch |

Source: Adapted from Department for Business, Enterprise and Regulatory Reform 2007, 16)

*Figure 3: The access criteria, detailed in Langford and Higgs (2010).*

## Summary of Literature Findings and Next Steps

The nature of financial exclusion is influenced by a range of social and spatial factors, some consistent and others nuanced, making it difficult for policymakers to fully understand and track (Devlin, 2005). Whilst cash use is falling and the distribution network shrinking, there is still a need to ensure people can access cash (Ceeney, 2019). The impact of bank branch closures has been felt disproportionately in deprived areas (Leyshon et al., 2008), and whilst the Post Office is attempting to fill this void, it is facing challenges in maintaining its network, with many consumers not viewing this as a credible solution to tackle financial exclusion (Midgley, 2005). Composite indicators have been successful in making complex issues easier to understand at the local level, for example in domains such as deprivation (MHCLG, 2019a), food deserts (Clarke et al., 2002) and loneliness (Lucy and Burns, 2017). However, work on financial exclusion indices has focused exclusively on the national scale, with little work to integrate the spatial distribution of infrastructure and consider local variations, something that this research will address.

## Methodology – Proposing a Framework

This section details the steps taken to develop a composite index to capture financial exclusion, and specifically to determine those areas at greatest risk.

This study focuses on the city of Nottingham, UK; however, the proposed index has been designed such that it can be readily applied to other areas where data are commonly available (primarily within the UK). Nottingham has been chosen as a case study city as despite being one of the UK's most deprived cities, its population is becoming younger and more diverse (McCurdy, 2019). As well as being home to the UK's first bank branch (Leighton-Boyce, 1958), the city has one of the highest levels of household debt of any UK local authority, with 1 in 20 residents having used a debt advice agency (Universities for Nottingham, 2020).

## Generating the Index

The index proposed in this research comprises three separate dimensions: [1] the ability to access cash (supply), [2] the distribution of people likely to be financially excluded and cash-reliant (demand) and [3] the alternatives available.

A shapefile (mappable boundary file) was obtained from the Open Geography Portal of the Nottingham City Council boundaries (boundaries for other cities can also be obtained in the same way, or from a central source such as the UK Data Service (2022)). This boundary file for Nottingham was then uploaded to DigiMap and 'Points of Interest' locational data obtained (see the categories and data processing required in Table 2). Provided by the Ordnance Survey, this free (for academic use) Points of Interest dataset contains the locations, addresses (where available) and mappable coordinates of over four million features of interest, such as retail premises, sports attractions and transport hubs. The data of all relevant cash infrastructure operating in September 2020 (most recent) were collected and then cleaned.

| OS CATEGORY | INFRASTRUCTURE | PRE-PROCESSING |
|---|---|---|
| 02,09,0138 | Banks and building societies | These data were cross-referenced with the individual bank's online branch locator tools and Which's list of bank branch closures since 2015 (Which?, 2020). Locations of bank contact centres or offices with no customer-facing facilities were removed, some of which were verified using Google Street View. |
| 02,09,0141 | Cash machines (ATMs) | These were cross-referenced with Link's online locator tool to establish whether they were free or fee-charging. If duplicates were identified, Google Street View was used to identify whether there were multiple ATMs at the same location (it was important for these to be retained as the loneliness of infrastructure would be analysed). Whilst this may skew the data by having abnormally high scores for one location, multiple ATMs would suggest strong demand for cash, high footfall and intensity of provision in this area. |
| 02,090811 | PayPoint locations | N/A. Not used in the original AvCash index. |
| 09,47,0667 | Frozen foods | All stores other than Iceland (which offers a cashback service) were removed. |
| 09,47,0699 | Convenience stores and independent retailers | Independent retailers were removed as it would be a huge undertaking to establish which offered cashback. The customer service teams of branded convenience stores (e.g. Londis) were contacted to establish their cashback policy and stores either removed or retained based on this outcome. |
| 09,47,0819 | Supermarkets | The customer service teams of supermarkets were contacted to establish their cashback policy, and stores either removed or retained based on this outcome. |
| 09,48,0763 | Post Offices | N/A |

*Table 2: Data categories downloaded from Ordnance Survey Points of Interest, relating to cash infrastructure*

The data were then sorted by postcode to identify instances where cash infrastructure may share the same location. For simplicity, locations with a free ATM at the same site as a branch or a Post Office were treated as two separate pieces of infrastructure. For major supermarkets such as Sainsbury's, Tesco and Waitrose, their policy is not to offer cashback if a free ATM is present, so the cashback provider was removed.

As reflected in Figure 4, the coordinates of each piece of infrastructure were plotted against the Lower-Layer Super Output Areas for Nottingham (a small geographic unit containing 400-1200 households, of which there are 182 units in Nottingham). The rationale for using LSOAs versus alternatives is detailed in a subsequent section of this paper. One example LSOA is shown in Figure 4, highlighted with the black outline. The population-weighted centroids for these LSOAs were downloaded from Open Geography Portal (2020c) (black point) and a catchment area buffer around them drawn. Multiple distances were assessed, but a distance of 500m was chosen (red circle), drawing largely on the literature on food deserts (Wrigley et al., 2004; Clarke et al., 2002) as an acceptable distance that someone may walk to access essential goods and services. Tischer et al. (2019) also considered 500m an acceptable distance to walk, although noted this should only apply to urban areas, of which all Nottingham's LSOAs are classified as. Whilst a network analysis was considered, this would increase complexity and many LSOA centroids do not fall directly on the road network. Further research could attempt to relate cash access to the transport network, particularly as many access cash as part of other journeys (Toynbee Hall, 2013), however, the index proposed in this research aimed to produce a simple and easy to replicate index. Although counting the infrastructure within the boundaries of each LSOA could produce useful results for large-scale analysis, it fails to reflect the reality that people move freely around into other LSOA's to access cash, it also results in ATM's within the reach of many LSOA's (in highly populated urban areas) only being counted once.

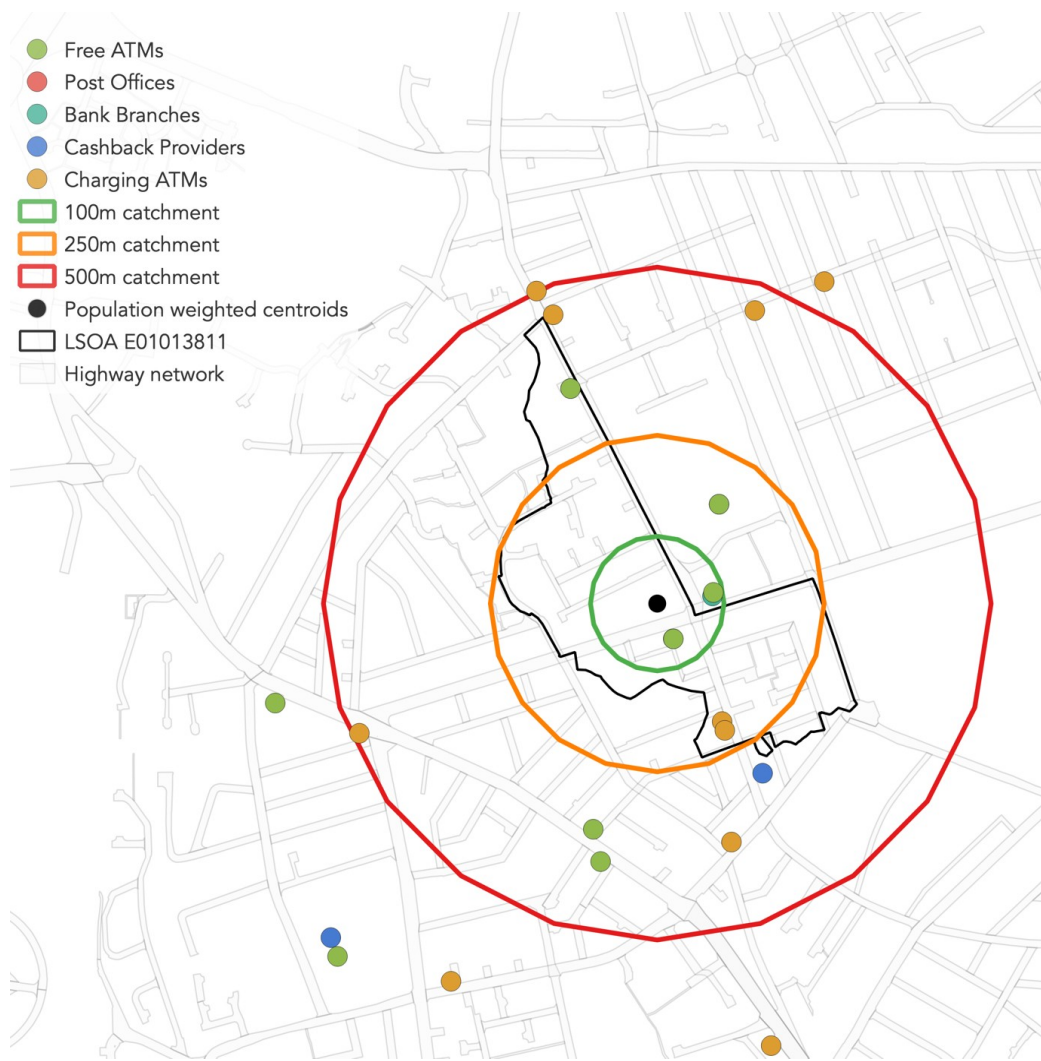

*Figure 4: Illustration of how the infrastructure within each LSOA catchment was collected.*

Each piece of infrastructure within the LSOA catchment area was then given a corresponding score, using the AvCash scoring system proposed in Tischer et al. (2019) and set out in Table 3. Whilst this is a subjective assessment (and some may place a higher value on accessing cash inside, others prefer the 24/7 convenience of an outdoor ATM), it is a compromise, with alternative scoring methods also discussed in Tischer et al. (2019). Modifying the scoring system was considered, however, upon reviewing the body of literature on cash use in the UK, the system proposed by Tischer. et al (2019) accurately reflected consumer preferences through qualitative research. The scoring system is based on the cost of accessing cash (with free methods deemed preferable to a charge or in return for another purchase) and the availability (based on opening hours and whether you need to be a customer of the institution to access cash, hence the Post Office scores higher than bank branches).

| Type of infrastructure | Score (per Unit) | Rationale |
|---|---|---|
| Free ATMs | 3 | Likely to be found in areas of high footfall with 24-hour access. As well as being free to use, these machines canbe accessed by any current account holder. |
| Post Offices | 2 | Withdrawals and deposits available to the vast majority of personal and business banking customers. Though opening hours are limited, services are free to use. |
| Bank/Building Society/Credit Union branches | 1 | Branch withdrawals are limited to the bank's own customers, and often have limited opening hours. However, they provide a free financial service that is highly desired amongst certain consumers. |
| Cashback providers | 0.5 | Free withdrawals, often in supermarkets, however this is largely dependent on making another purchase. Some providers may place a cap on withdrawals |
| Fee-charging ATMs | -0.5 | These machines charge to withdraw money, often located within newsagents or convenience stores which can make access difficult, though they are the only option for many consumers. |

*Table 3: Assigned availability of cash scores per unit of infrastructure, as detailed in Tischer et al. (2019)*

An example of using this scoring system (Table 3) to provide numeric scores for LSOA's is presented in Table 4.

| LSOA | Free ATMs | POs | Branches | Cashback | Charging ATMs | Total Score |
|---|---|---|---|---|---|---|
| E01013941 | 1 | 0 | 0 | 1 | 1 | 3 |
| E01013957 | 1 | 1 | 0 | 2 | 1 | 5.5 |
| E01013838 | 3 | 0 | 1 | 1 | 0 | 10.5 |

*Table 4: Example of LSOA scoring*

The next stage of the research was to identify areas that could lose access to cash, because an ATM malfunctions, runs out of banknotes or is removed completely as its unprofitable. 'Lonely' ATM's are identified where there is no alternative within 250 metres, thus causing inconvenience should it be removed from the system. Supermarkets and Post Offices were omitted from this analysis as they also have alternative uses, meanwhile bank branches must go through a consultation process before closure and tend to leave ATM's in situ if a branch closes. This analysis was completed using the 'join attributes by nearest function' in QGIS, an opensource and freely available GIS package, to calculate the distance from the nearest neighbour. The data were then sorted by distance, identifying machines with alternatives within 100, 250, 500 and over 500 metres. ATM's with no alternatives within 250 metres were mapped against the original LSOA centroid catchment areas, and the number of 'lonely' ATMs within each catchment area were counted.

# Financial Exclusion

The next stage in the process involved ascertaining the variables that identify areas with potentially high levels of financial exclusion, based on the characteristics defining financially excluded population groups. Whilst there has been some statistical work to establish the indicators of financial exclusion (Devlin, 2005; Hogarth and O'Donnell, 2000; Kempson et al., 2000; Select Committee on Financial Exclusion, 2017), there has been very little work to quantify this at the local scale. Upon reviewing the literature, many factors were found to influence financial exclusion, and these are detailed in Table 1, however, only the most pronounced and frequently mentioned factors are carried forward for use in the index, and these are summarised in Table 5, together with firm rationale. The index was constructed using a common additive method, widely used in other domains. The approach is outlined in Gibson and See (2006).

| Variable | Indicator | Rationale for inclusion |
|---|---|---|
| **Affecting Supply** | | |
| **Availability of infrastructure** | AvCash Score | Provides a useful indicator of accessibility and availability of cash. |
| **Loneliness** | Number of ATMs with no alternatives within 250m | High risk of areas losing access temporarily through malfunctions, or permanently through withdrawal. |
| **Affecting Demand** | | |
| **Employment status** | Claimant count as a proportion of LSOA population | "*Consistent and marked influence*" (p.96) on financial exclusion across all products (Devlin, 2005). |
| **Income** | Admin based income statistics: Combining PAYE and benefits payments to estimate net income at LSOA level | Appeared to be the strongest and most widely recognised indicator, with strong links to the poverty premium (Osborne, 2015). |
| **Housing tenure** | Number in private rented, social rented or part-owned accommodation as a proportion of households in an LSOA | A consistent indicator, with strong risks of emerging pockets of exclusion in social housing. |
| **Lone parents** | The proportion of households in an LSOA headed by a lone parent | Very low use of financial services, with a risk of deep exclusion, but not as pronounced as other variables. |
| **Affecting Access to Alternatives** | | |
| **Ability to access banking online** | Internet User Classification | LSOAs are assigned a score, 1 being the committed and regular users of the internet, 10 being the least, based on the population's behavioural characteristics and internet shopping habits (Alexiou and Singleton, 2018). The data were reversed (so a score of 1 would become 10) and standardised to fit within the index. |
| **Ability to travel to access cash** | The proportion of households with an LSOA with access to a car | Increased vulnerability if individuals are not able to travel to access alternatives. |

*Table 5: Index variables, indicators, reasoning and weights*

The data were also designed to be as easily obtainable as possible, using only freely publicly accessible data. Whilst efforts were made to source the most recent (and regularly updated) variables, census data for some variables were needed, notwithstanding that this is now close to ten years old (2011 Census). Alternative options could include ONS statistical estimates, however many of these are at middle-layer super output areas (MSOA) levels. The release of the 2021 Census data represents an opportunity to update the index results.

When selecting the scale for analysis, it was important to consider the Modifiable Areal Unit Problem, a form of bias in spatially aggregated data, common in choropleth mapping (Openshaw, 1981). The LSOA unit was chosen as the spatial scale, not only as there is a wide quantity of data available at this scale, but also it matches the original AvCash index developed by Tischer et al. (2019), allowing for comparison with the analysis undertaken on Bristol. Additionally, the scale is small enough to limit the ecological fallacy but large enough to identify neighbourhood effects and shared characteristics of households. This unit of analysis can also be used to compare the average cash score against other classifications (or indices) designed to fit census boundaries. Census wards were also considered as a scale of analysis, as they may show some wider patterns that may be lost at the finer LSOA level, with areas being more familiar and identifiable by name, although they are larger than MSOAs which had been discounted due to their size not lending themselves to detailed analysis. As an urban unitary authority, Nottingham's large wards tend to group areas with differing characteristics (such as high-income Wollaton and relatively deprived Lenton), and the boundary changes in 2019 means there is only a limited amount of data available using the new boundaries. Although the data wasn't mapped at the ward level, the boundaries were overlaid to aid understanding.

The data were then pre-processed, firstly to ensure that they all had the same polarity (data direction). In this index, a higher score would indicate a greater extent of financial inclusion, meaning much of the data had to be reversed (for percentages, this was simply a case of subtracting 100 from the value, as shown shortly). Analysis was carried out to identify any correlation between the datasets. Openshaw (1995) suggests a correlation coefficient of over 0.95 is grounds to exclude a variable as the relevant dimension is likely captured more than once. The only variable pair indicating multicollinearity (0.9) was car/van availability and housing tenure, however, there is little correlation between these two variables and other variables within the analysis. Furthermore, as the literature highlights the importance of these variables in establishing financial exclusion, they remain in the index. The data were then standardised, with raw variables being converted to a percentage using either a population or household count denominator, and normalised using the Min:Max approach.

The Min:Max approach operates whereby $x_{raw}$ is the raw variable, $min_i$ is the minimum value for each variable, whilst $max_i$ is the maximum value, i is the variable number from 1 to 0. This easy to implement scaling method rescales the data in a range from 0 to 1, allowing normalisation to take place across varying data types (e.g. raw counts, percentages and distance variables).

$$x_{norm} = (x_{raw} - min_i)/(max_i - min_i)$$

The next stage was to consider weights for the variables within the index. The first approach considered was not to assign any weights (and in effect employing blanket equal weighting). When attempted, this approach was rejected, primarily for three reasons:

- Several areas with extremely low access to cash scores ultimately scored highly for financial inclusion.
- The index focused too heavily on demographics factors, which risked creating a deprivation index with little exploration of infrastructure provision.
- District centres (such as retail centres) with high cash provision scored poorly.

The weightings were then re-evaluated, with the three domains of supply, demand and alternatives awarded equal importance (33.3%). The sub-variables were then weighted as shown in Table 6, using intelligence garnered from academic literature (Table 5), in addition to Nottingham-specific findings.

| Supply | Demand | | | | | Alternatives | |
|---|---|---|---|---|---|---|---|
| Availability of Infrastructure | Loneliness | Claimant Count | Income | Housing | Lone Parents | Internet User Classification | Car Tenure Availability |
| 26.67% | 6.67% | 9.52% | 9.52% | 9.52% | 4.76% | 16.67% | 16.67% |

*Table 6: Variable weightings*

The financial exclusion index was then mapped against the LSOA's using QGIS and presented in five categories using Jenks' Natural Breaks distribution. This brought the benefit of placing values in naturally occurring data categories, with breaks between the categories maximised and variation within categories minimised, allowing a choropleth map that represents true trends in the data (Map 1). The index score for each LSOA was then compared with the LSOA Output Area Classification [OAC] (ONS, 2021a) for that particular area. These areas were mapped using QGIS and represented in Figure 5, detailing the financial exclusion index score, percentage of LSOA's and details of average characteristics for certain factors of these areas in Nottingham. Whilst these factors do not necessarily influence financial exclusion, they give a useful indication as to the population and socioeconomic characteristics of these areas and can make financial exclusion more accessible to policymakers who may be well versed in such classifications.

To fully understand the nature of cash infrastructure, a thematic map was created by joining the shapefile of Nottingham's LSOA boundaries to the AvCash score for each area (shown in Map 2). To identify the geographical variation of various types of infrastructure, a thematic map was created for each infrastructure type (Free ATMs, Post Offices, banks, supermarkets and charging ATMs), with the number of units accessible from each LSOA overlaid with the individual points of each unit (see Maps 3-7). A nearest neighbour analysis was conducted using QGIS' Nearest Neighbour Analysis tool, to understand whether infrastructure is evenly distributed across the city, or clustered around certain areas, the results of which can be found in Table 7.

To validate the index, the scores for all LSOA's were then compared with the rankings for wards produced by Experian in 2007 (Nottingham Financial Resilience Partnership, 2020), where UK census wards were ranked from 1-10,000 on their levels of financial exclusion. Within Nottingham, 95% of wards fell in the top two septiles for financial exclusion. Although the methodology used to create the Experian index has not been published, this approach highlights if there is any consistency in between findings. Given the difference in scale, methodology and data, no attempt will be made to analyse the outcomes as there are likely to be significant statistical errors. This research will only seek to determine if there is a correlation between the rankings and any noteworthy changes in the seventeen-year period. To complete this, an LSOA to ward lookup file was downloaded from the ONS' Open Geography Portal, filtered to only included Nottingham (using pre-2019 boundaries). The median index scores of the LSOAs within each ward was calculated and then ranked from 1 to 20. This was compared with the rankings for the 2007 study and the differences between the rankings calculated to identify any significant changes. The differences in rankings were also mapped on a choropleth map using QGIS (Map 8).

# Patterns of Financial Exclusion

The index set out above was then implemented and ultimately visualised using QGIS, an open source and freely available mapping and spatial analysis software. The outcome can be seen in Map 1. Map 1 shows a high level of financial inclusion in affluent Wollaton to the west, certain city centre LSOA's, as well as in The Park (a private residential estate), Wilford Village and Mapperley Park. The high scores of the city centre are not surprising, with many financial infrastructures spatially concentrated, though some LSOA's score lower as their population-weighted centroids are further away from the infrastructure-dense centre. The high concentration of students and halls of residences in the neighbouring LSOA's to the south and north of the city centre may distort the income statistics (students generally don't pay tax or receive benefits), giving a somewhat distorted impression of deprivation. Whilst the higher scoring LSOA's tend to be more affluent areas of the city, other relatively deprived areas such as The Meadows, Dunkirk and Rise Park score well, indicating a high level of infrastructure provision despite their relative deprivation. Despite Compton Acres' AvCash score of 0.5, it sits comfortably in the fourth quintile for financial inclusion, being the LSOA with the highest median income at £15,909. Most notably, there are pockets of high financial exclusion concentrated in Aspley, Bestwood Park estate, Hyson Green, Radford and St. Ann's, broadly consistent with the findings of Nottingham Financial Resilience Partnership (2020).

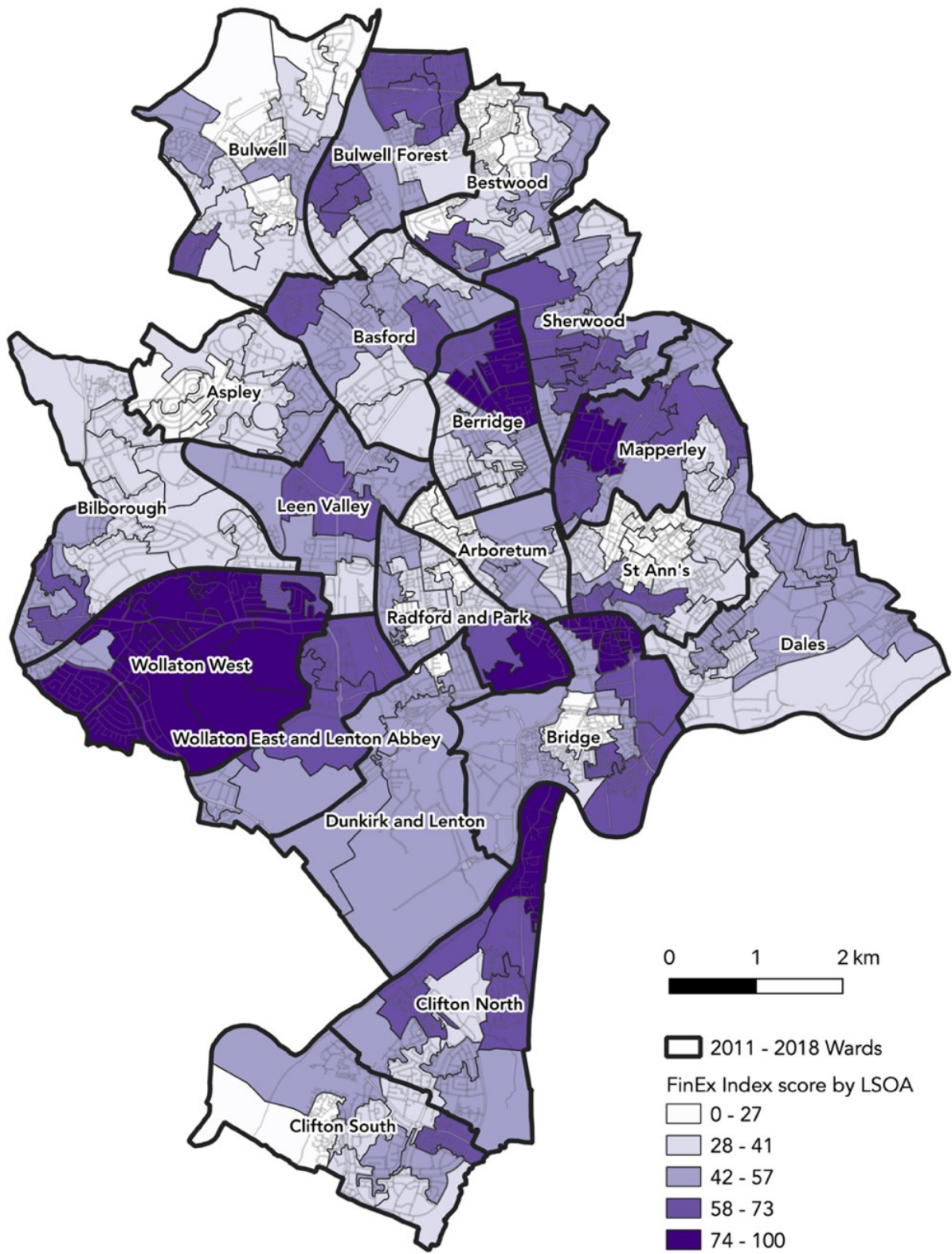

*Map 1: Financial exclusion index scores (higher score indicates a greater level (risk) of financial exclusion)*

When contrasting the financial exclusion index with the OAC, for both corroboration and validation purposes (see Figure 5), high financial inclusion areas are found in LSOA's categorised as 'Suburban Living' (Wollaton, Wilford and parts of Bestwood), defined by low-density and owner-occupied housing, low unemployment and an ageing population (ONS, 2018). Residents are more likely to use private transport (important given the relatively low level of infrastructure within these areas), though these areas comprise only 4.9% of Nottingham's LSOAs.

The lowest financial exclusion scores are found in areas classified as 'Hard-Pressed Communities', making up over one quarter of Nottingham's LSOAs, predominantly found in locations such as Clifton, Bulwell, Aspley and Bestwood. As well as being characterised by a primarily white population, housing is largely terraced with high proportions of social renting and above-average unemployment. These areas are almost exclusively former council estates and despite improvements to the fabric of the housing stock (Nottingham City Council, 2020), there are noteworthy problems of housing insecurity, strained household budgets and many on low incomes reliant on state support to survive (Nottingham City Council, 2013). The areas often lack mainstream cashback providers (indicating a wider issue of food deserts (Clarke et al., 2002)), 82% of ATM provision is by private, independent ATM deployers (IADs) such as Cardtronics, which operate in a different way to banks, and more likely to close machines due to declining usage.

## Cosmopolitan Student Neighbourhoods

15.4% of LSOAs

**61.8**
Median Financial Exclusion Index Score

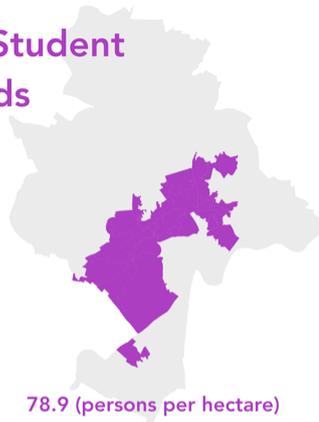

| | |
|---|---|
| Population density | 78.9 (persons per hectare) |
| Median age | 24.2 |
| Claimant count | 3.7% |
| Households without a car | 53.6% |
| Households owner occupied | 25.6% |
| Student population | 55.4% |
| White population | 64.8% |

## Ethnically Diverse Professionals

15.9% of LSOAs

**62.4**
Median Financial Exclusion Index Score

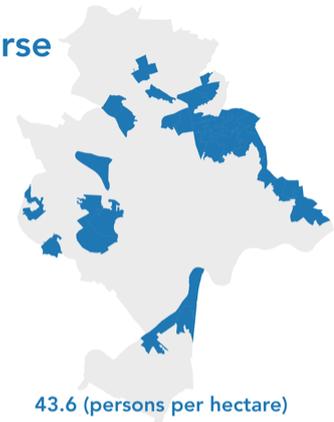

| | |
|---|---|
| Population density | 43.6 (persons per hectare) |
| Median age | 39.5 |
| Claimant count | 4% |
| Households without a car | 26.7% |
| Households owner occupied | 70.2% |
| Student population | 7.4% |
| White population | 78.5% |

## Hard-Pressed Communities

25.8% of LSOAs

**38.1**
Median Financial Exclusion Index Score

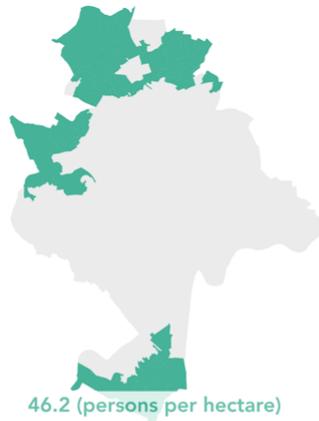

| | |
|---|---|
| Population density | 46.2 (persons per hectare) |
| Median age | 36.6 |
| Claimant count | 6.3% |
| Households without a car | 46.5% |
| Households owner occupied | 47.5% |
| Student population | 6.9% |
| White population | 84.7% |

## Industrious Communities

2.2% of LSOAs

**46.2**
Median Financial Exclusion Index Score

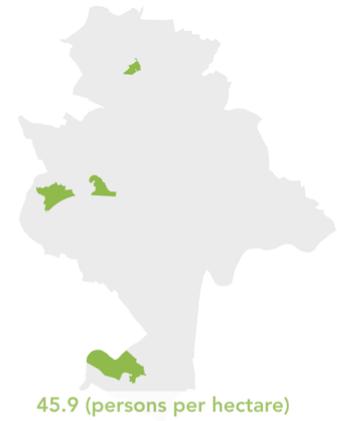

| | |
|---|---|
| Population density | 45.9 (persons per hectare) |
| Median age | 44.4 |
| Claimant count | 3.8% |
| Households without a car | 33.1% |
| Households owner occupied | 64.4% |
| Student population | 5.4% |
| White population | 88.9% |

## Multicultural Living

35.7% of LSOAs

**41.4**
Median Financial Exclusion Index Score

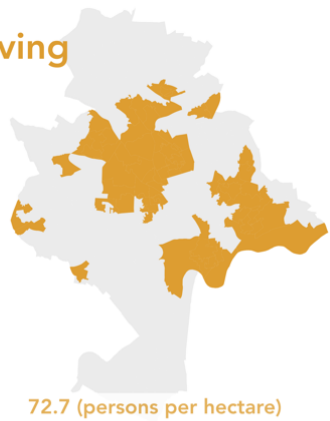

| | |
|---|---|
| Population density | 72.7 (persons per hectare) |
| Median age | 31.8 |
| Claimant count | 7% |
| Households without a car | 49.1% |
| Households owner occupied | 39.4% |
| Student population | 13.1% |
| White population | 60.4% |

## Suburban Living

4.9% of LSOAs

**70.1**
Median Financial Exclusion Index Score

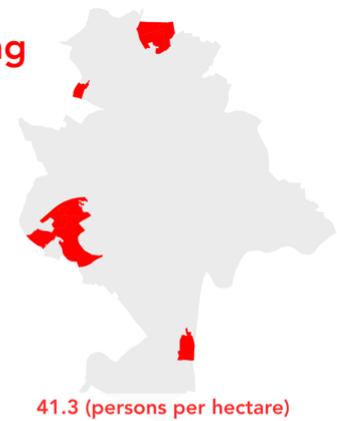

| | |
|---|---|
| Population density | 41.3 (persons per hectare) |
| Median age | 47.8 |
| Claimant count | 2.2% |
| Households without a car | 14.2% |
| Households owner occupied | 89.9% |
| Student population | 6.3% |
| White population | 82.7% |

*Figure 5: Contrasting the Financial Exclusion Index with the OAC*

# Similar Index Scores, Different Ground Conditions

Whilst additive composite indices have their benefits, not least a simple and understandable workflow, it can be the case that certain areas have very similar scores but very different underlying characteristics – something which the index fails to accurately capture.

An example of this can be seen when comparing two LSOA's, one in Arboretum (scoring 24.24) and the other in Bestwood (scoring 24.42) (see Figure 6), both in the lower quartile of index scores. Despite the LSOA in Arboretum having high levels of cash infrastructure, there are many charging ATM's. Most ATM's are based at convenience stores, and 75% operated by IADs (whereas Bestwood is entirely provided by IADs), creating an increased risk of withdrawal – private operators may not be as concerned with the social purpose of providing cash. If infrastructure were to be withdrawn, the predominant Internet User Classification (IUC) is of 'Passive and Uncommitted Users', whilst 32% own a car. Bestwood's cash provision is considerably poorer. Whilst more own a car (50%), the LSOA is in the bottom category for internet usage, and it is a long distance to the nearest bank branch. The LSOA is only slightly more affluent than that in Arboretum across all index categories. This demonstrates the usefulness of the index in showing the risk of financial exclusion – despite Arboretum's good infrastructure provision, the heightened risk of withdrawal and more geodemographics that indicate financial exclusion puts it on a similar level to Bestwood which is already sparse in infrastructure.

Similar differences can also be seen when contrasting two other LSOA's in the median quartile, in St. Ann's (scoring 50.08) and Leen Valley (scoring 50.34) (see Figure 7). Despite an LSOA in St. Ann's being in the second IMD decile (MHCLG, 2019b), the western edge is in easy reach of much of the city centre's ATM's (though only 20% are bank-owned), however, provision closer to the centre of St Ann's and the east is, foremost, charging ATMs. Over half of households rent from a housing association, and another 30% privately, characterised by tower blocks and extensive student accommodation. Whilst the predominant IUC is the 'Youthful Urban Fringe', 69% own a car. The Leen Valley LSOA is three miles from the city centre, characterised by low-density housing and car ownership of 73%. The claimant count is low with a median income of £12,213 (0.73 on the standardised scale). However, charging ATMs are plugging gaps in supply, with the nearest free alternative being over 850m away, and an AvCash score of -0.5. The neighbourhood is also profiled as one of passive and uncommitted internet usage, suggesting challenges in meeting financial needs.

LOWER QUARTILE

LSOA E01013811

Arboretum

*Score: 24.24*

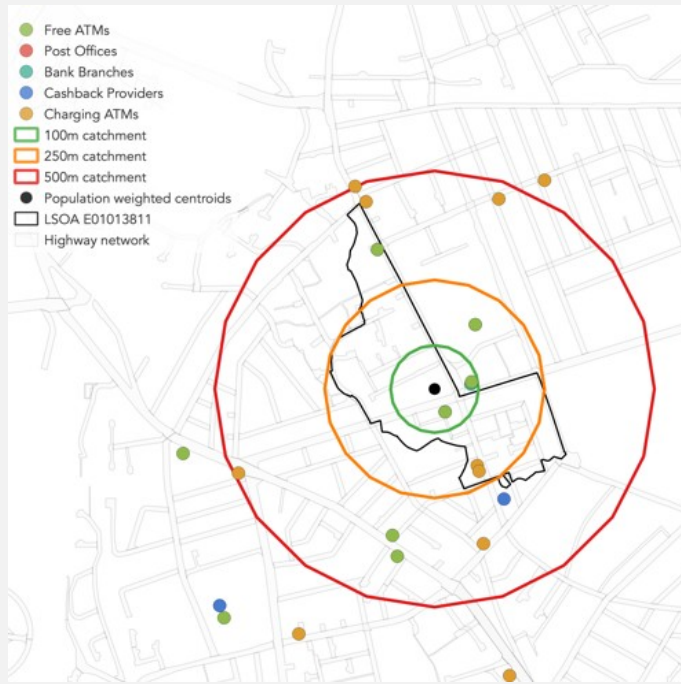

LSOA E01013850

Bestwood

*Score: 24.42*

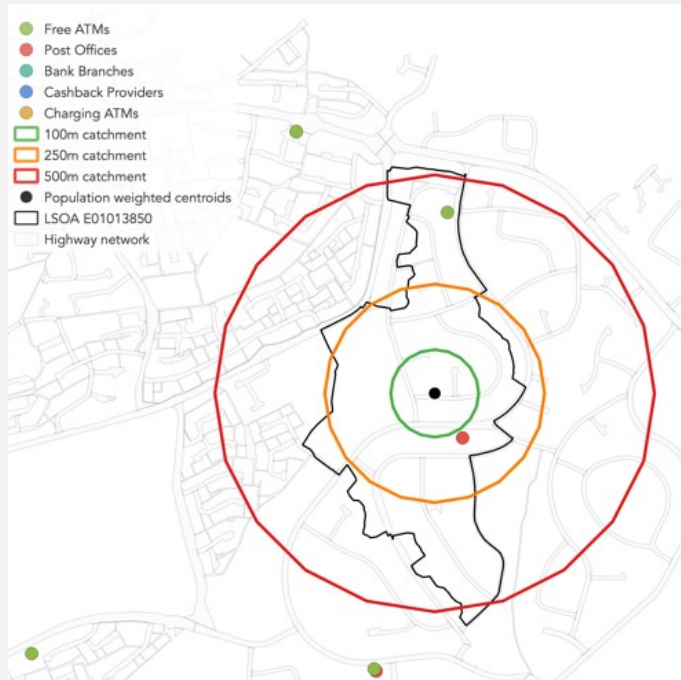

*Figure 6: Contrasting LSOA's with similar scores in lower quartile.*

MEDIAN

LSOA E01033399

St. Ann's

*Score: 50.08*

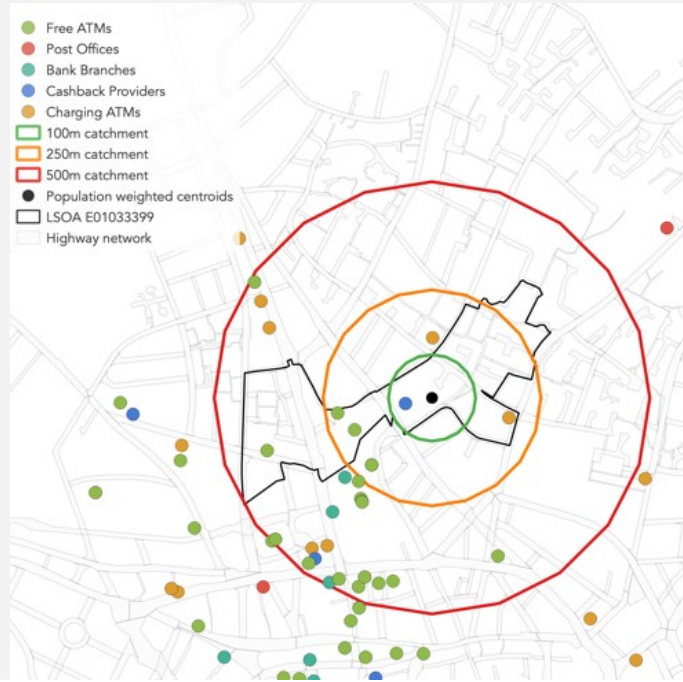

LSOA E01013932

Leen Valley

*Score: 50.34*

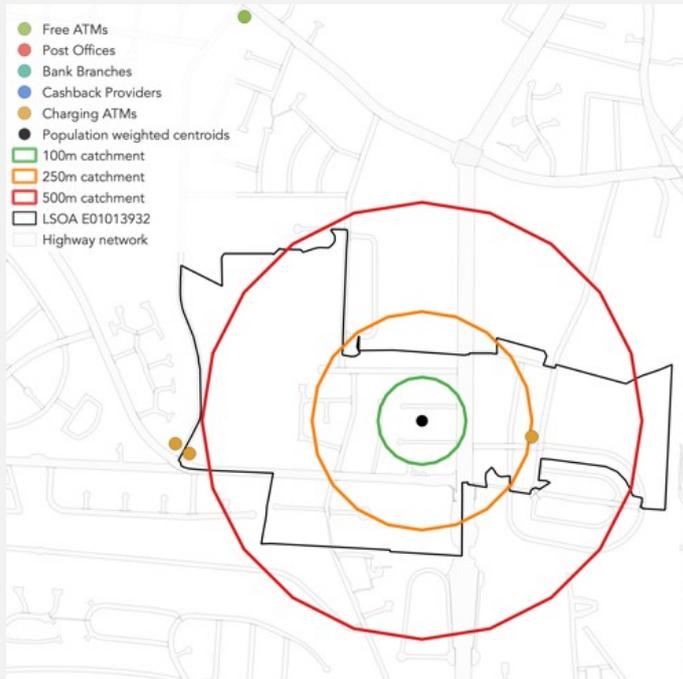

*Figure 7: Contrasting LSOA's with similar scores in median quartile.*

# The Geographical Distribution of Financial Infrastructure

Map 2 shows the standardised AvCash score by LSOA, (previously set out in Table 3), whilst Table 7 demonstrates the distribution of cash infrastructure through a nearest neighbour analysis. The findings are consistent with French et al. (2008) and Tischer et al. (2019), with infrastructure generally concentrated in city and district centres and more affluent neighbourhoods.

| Infrastructure | Number of points | Expected mean distance (m) | Observed mean distance (m) | Nearest neighbour index | Z-score | Interpretation |
|---|---|---|---|---|---|---|
| Free ATMs | 225 | 353.1 | 145.7 | 0.4 | -16.85 | Very clustered |
| Post Offices | 30 | 925.6 | 1032.9 | 1.1 | 1.21 | More regular |
| Bank/building society branches | 33 | 689.5 | 526.4 | 0.8 | -2.6 | Quite clustered |
| Cashback providers | 42 | 753.5 | 599.6 | 0.8 | -2.53 | Quite clustered |
| Charging ATMs | 83 | 540.3 | 377.6 | 0.7 | -5.24 | Quite clustered |

*Table 7: Nearest neighbour analysis for cash infrastructure in Nottingham*

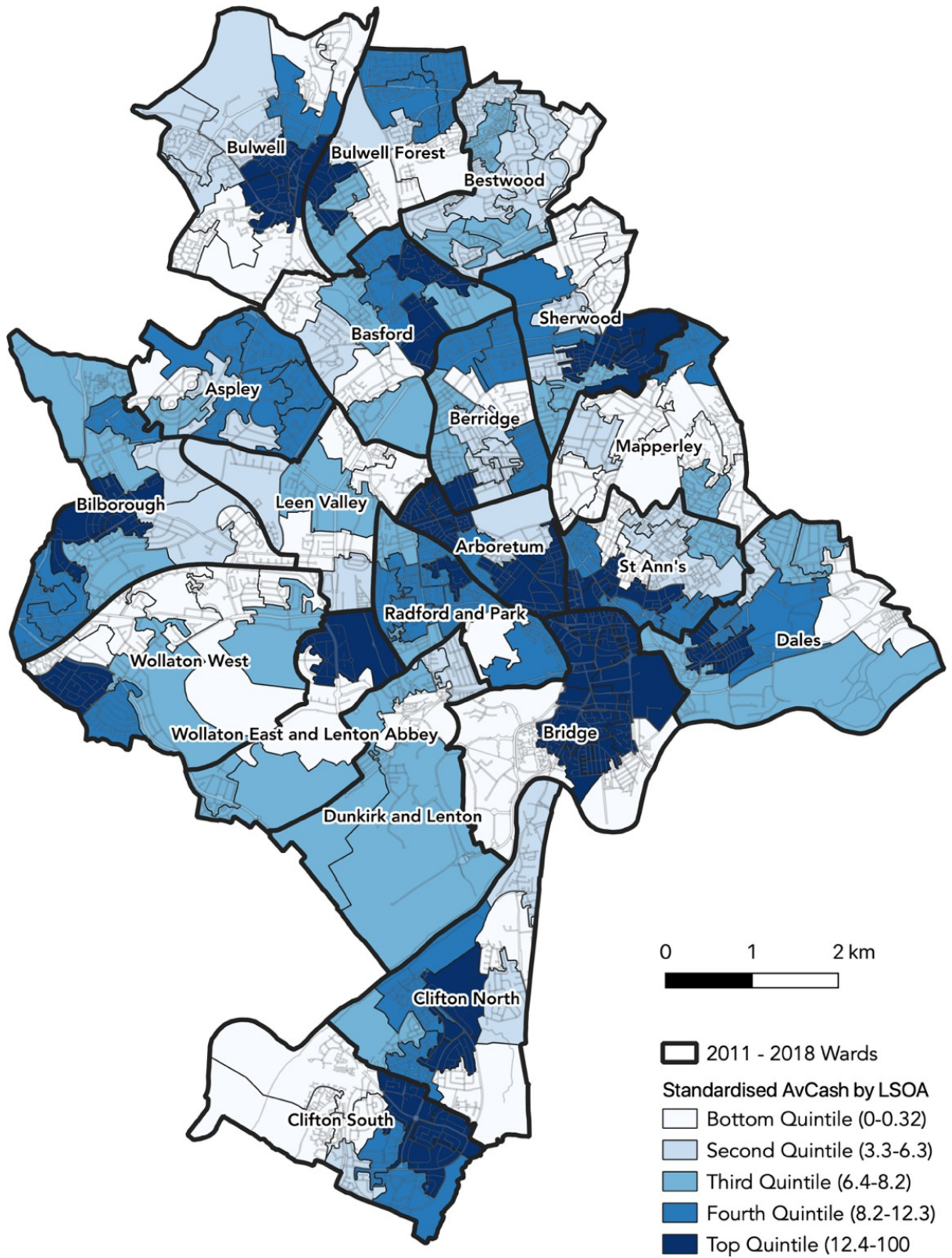

Map 2: Availability of Cash (AvCash) Score by LSOA

## FREE ATMS

The dominant infrastructure, with 225 machines making up 44% of access points, with the city centre and district centres very well catered for, whilst the suburbs are generally well served with 82% of LSOAs being within 500m of a free ATM, though the residential areas of Basford, Beechdale and Bestwood plus industrial areas appear to have a lower level of access.

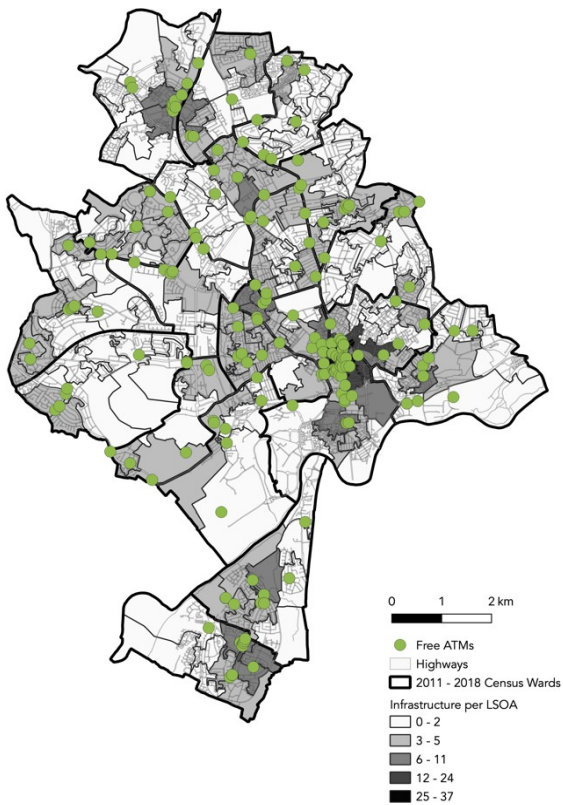

*Map 3 (above, left): Distribution of Free ATMs*

*Map 4 (below, left): Distribution of Post Offices*

## POST OFFICES

Post Offices have a more regular distribution across the city (see Table 7), largely because of the network access programme setting standards (Langford and Higgs, 2010). Despite only 5% of cash users withdrawing regularly from the Post Office (Parrott, 2018), their distribution means they are in a better position to provide financial services than many branches. Though they're not immune to wider societal changes, with the shift to online and low transaction fees for banking meaning some are planning to close (Monaghan, 2019). The National Federation of Sub Postmasters has called on the government to mandate that banks fund a free-to-use ATM network and Post Office services with the savings gained by closing branches (Hall, 2020).

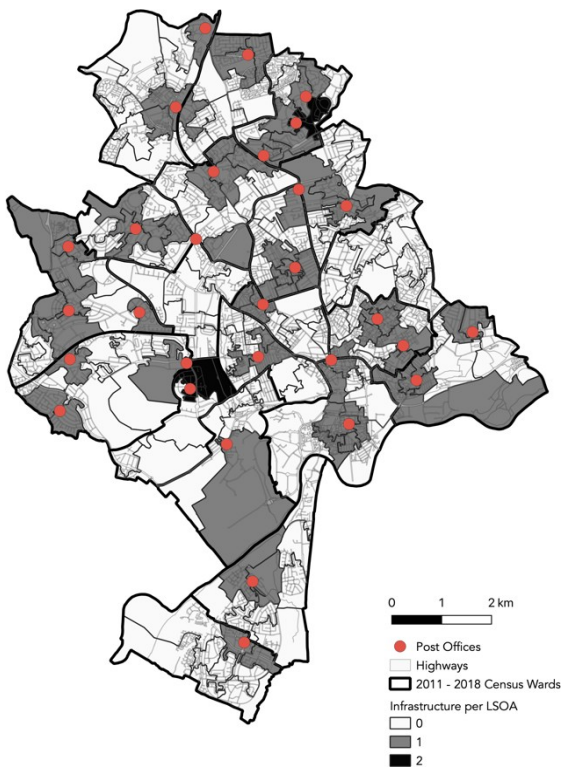

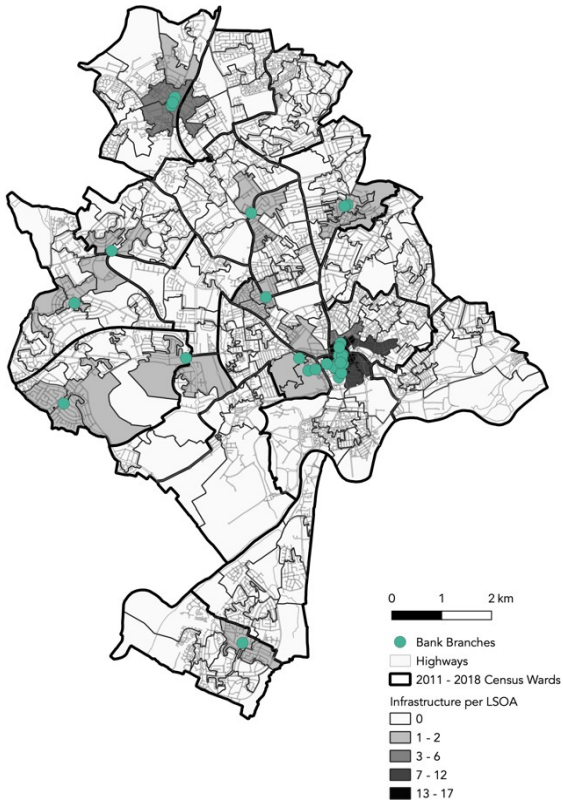

*Map 5 (above): Distribution of bank branches*

## BANK BRANCHES

33 branches serve Nottingham, though a 58% fall from 2015 (Which?, 2020), heavily concentrated in the city and some district centres. The south and east of the city appear to be poorly served, though this analysis does not include branches in towns over the border in Carlton and West Bridgford, which many residents will access. Nottingham's branches do not seem as clustered as Bristol's (Tischer et al., 2019), with many in the west of the city maintaining relatively even coverage, though 55% of LSOA centroids are over 1km away from their nearest branch. As branches are exclusive to their customers, many will live close to a branch but must travel into the city centre to access services. Of the 19 banking brands, only four serve customers in the suburbs.

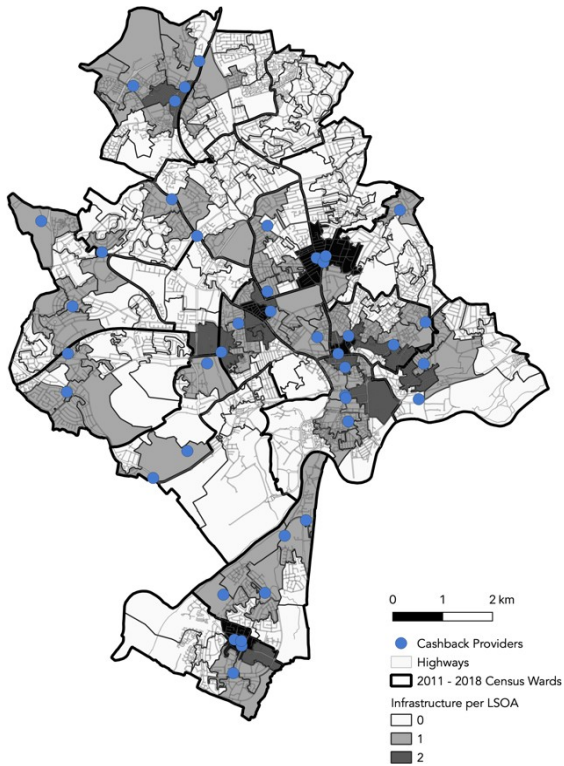

## CASHBACK PROVIDERS

Cashback providers are akin to branches in having a nearest neighbour score of 0.8, being spatially clustered in areas of economic activity. There are notable gaps in provision, particularly around Bestwood, Leen Valley and Lenton, though provision may be higher if small independent retailers were mapped (see Table 2).

*Map 6 (above, left): Distribution of cashback providers*

*Map 7 (below, left): Distribution of Charging ATMs*

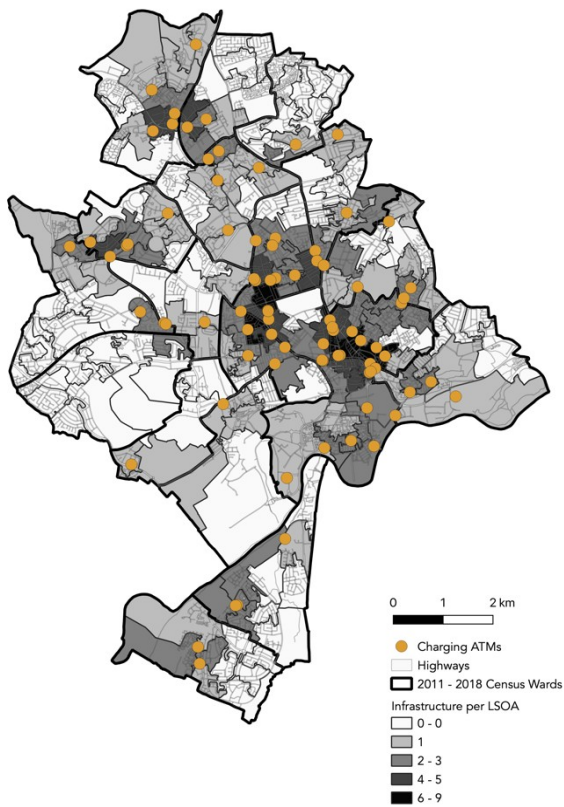

## CHARGING ATMS

Charging ATMs appear to plug some of the gaps in the provision of free ATMs. Unlike in Tischer et al. (2019), there are few ATMs in the city centre, but strong provision to the north of the centre, and in the areas of Radford, Lenton and Hyson Green. The LSOAs with the highest number of charging ATMs tend to be in cosmopolitan student neighbourhoods as well as in petrol stations, industrial areas and within pubs, away from residential areas but often the only way to access cash. Similar to Tischer et al. (2019), there is an absence of charging ATMs in affluent wards, such as Wollaton and The Park.

# The Challenge of Isolated ATM's

Even in areas with good coverage, access can be hampered if an ATM malfunctions, runs out of cash or is removed completely if unprofitable. Whilst LINK has a scheme to protect ATMs in low-income areas with no alternative within 1km, Evans et al. (2020) found this distance too far for many, particularly those with mobility impairments or a health condition. Figure 9 shows that 22% of Nottingham's ATMs' are over 250m away from their nearest neighbour, compared to 25% in Bristol (Tischer et al., 2019). Charging ATM's are far less likely to have an alternative ATM nearby when compared to free ATM's (with 55% being over 250m from an alternative), which fits the narrative of charging machines plugging gaps in supply. When considering the nearest free machine, Figure 10 shows that 29% of all machines have no alternatives within 250m, whilst 49% of all charging ATMs have no free alternative within 250m, illustrating significant barriers to free cash access in some LSOAs.

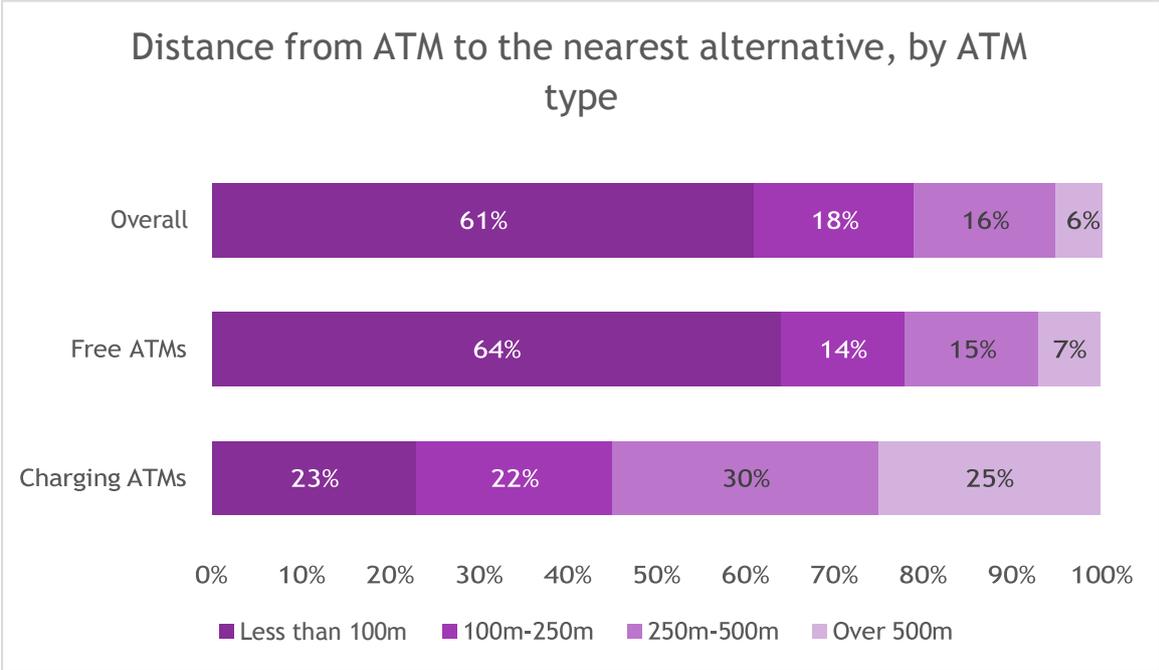

*Figure 9: Distance from ATM to the nearest alternative*

Whilst the Citizens Advice Bureau (2006) and Tischer et al. (2019) find charging ATM's concentrated in areas of deprivation, this research does not come to the same conclusion. Despite high concentrations of charging ATM's in some deprived areas (notably Aspley, Hyson Green and St. Ann's), it is not a trend across Nottingham and statistical tests cannot find a correlation between the Index of Multiple Deprivation and the number of charging ATMs.

Further work could result in a longitudinal study, analysing whether there are an increasing number of ATM's being converted to charging machines and whether these are concentrated in deprived areas.

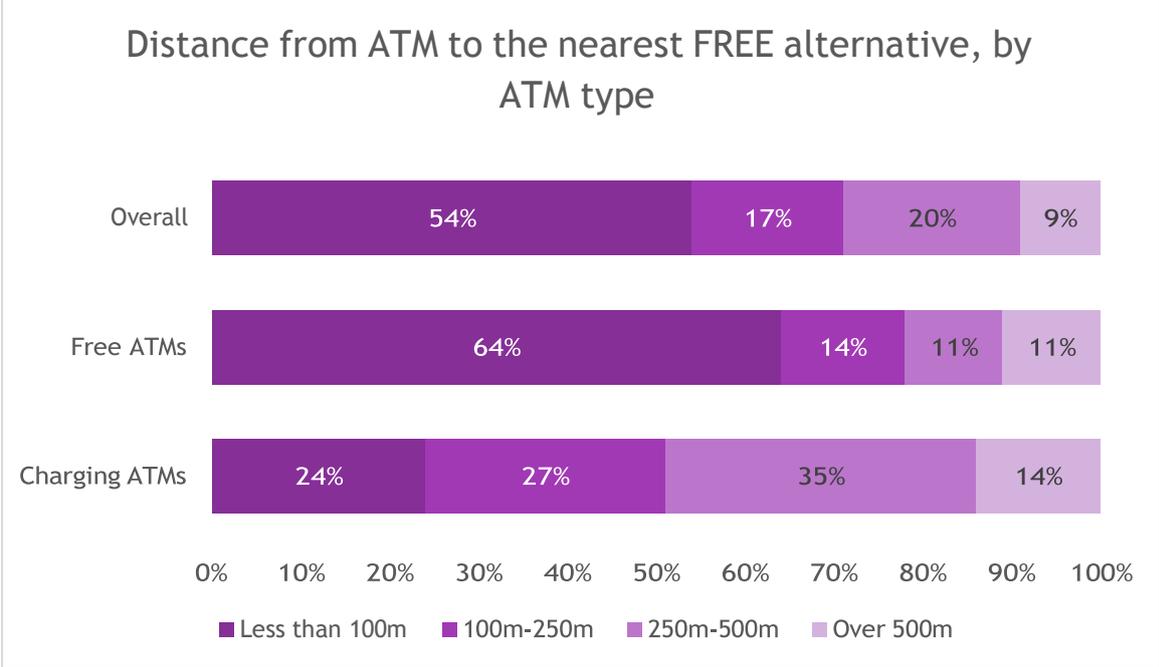

*Figure 10: Distance from ATM to the nearest free alternative*

# Validating the index

The Nottingham Financial Resilience Partnership (2020) is a steering group of public and third-sector organisations aimed at improving financial resilience and access. It works on a hub and spoke model, with local communities feeding into a wider strategic group. The four priority areas are Bestwood, St Ann's & Sneinton, Clifton & Meadows and Aspley. The financial exclusion index confirms that Aspley and St. Ann's are key pockets of financial exclusion, with the second and third-lowest median index score respectively. Whilst The Meadows is relatively deprived, it benefits from a good level of cash access and easy access to local centres with high levels of financial service provision. It is also undergoing a radical redevelopment with a growing influx of students and young professionals, possibly distorting the income statistics. There is variation in Clifton too, the northern LSOA's score well, though the southwest of the estate, with the isolated Clifton Village and areas around the A453, score poorly.

Bestwood tends to score poorly for financial inclusion, though there is much variation in the north of the city. Whilst the four priority areas seem to fit with the indexes finding, it would make more sense to introduce Bulwell as an additional priority area and reconsider the focus on Clifton.

Experian produced a 2007 report (Nottingham Financial Resilience Partnership, 2020), ranking census wards on their levels of financial exclusion. Given possible statistical errors outlined earlier, it would be unwise to draw detailed conclusions from this dataset, however, there is a correlation of 88% when comparing the ward rankings between 2007 and 2020. Whilst not significant, it demonstrates there are broad similarities. Map 8 demonstrates changes over time, with Bridge ward appearing to have escalated seven places. This would appear to be an outlier, given it includes the infrastructure-heavy city centre and there have been significant socio-economic changes within this ward. It also suggests Berridge ward has fallen five places, which could be attributed to the rising student population and high deprivation to the south, though it is also an area of growing affluence to the north. Generally, wards that have seen extensive economic development (such as Lenton, around the universitycampuses), have improved whilst some inner-city wards are falling down the rankings. Though it is challenging to make assumptions when the variables in the Experian index are unknown, there is some correlation between the indices.

Validation is made more difficult as the concept and determinants of financial exclusion are themselves contested. In recent years, attention has turned to the role of ATM's and internet usage in determining how people access the financial system (Nieboer, 2019), other research has focused on bank account ownership, people's propensity to save or insurance take-up (Collard, 2007). However, people would need to hold a bank account to use an ATM or Post Office, and the likelihood of owning an account (for cash-heavy users) is determined by infrastructure provision. Whilst this index has attempted to relate infrastructure provision to the determinants of financial exclusion, further research is needed on the strength of that relationship.

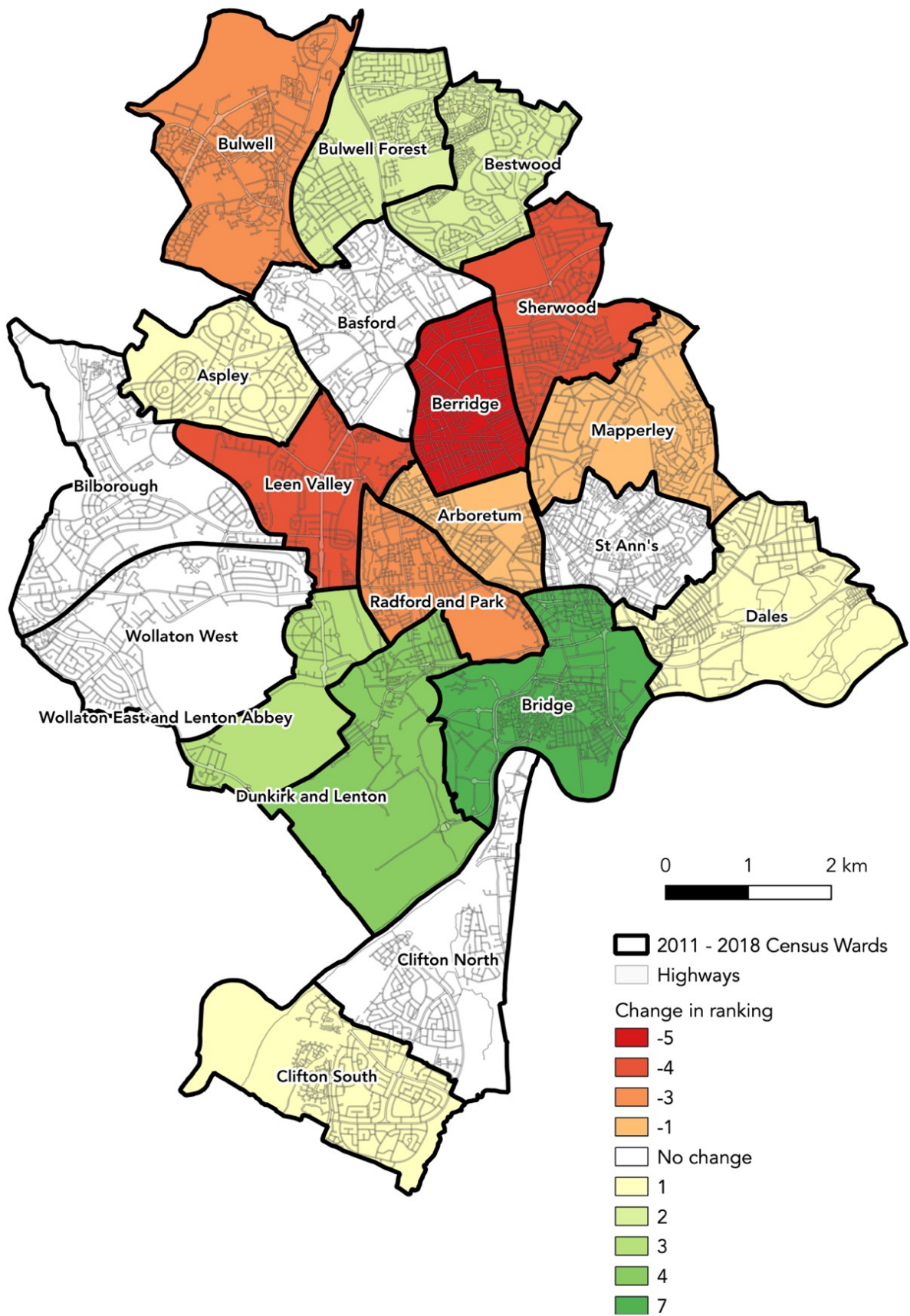

*Map 8: Change in ranking from the 2007 Experian index*

# Implications for Policy Makers

The index shows that financial exclusion is complex and nuanced, influenced by a variety of factors, with areas scoring similarly demonstrating varying characteristics. It provides an opportunity for organisations to identify areas at risk of exclusion and examine how infrastructure withdrawals or wider social / economic changes could affect this.

## Understanding How Infrastructure Withdrawal Affects Financial Exclusion

Despite six bank brands recently announcing closures, only one (TSB's city centre branch) in Nottingham will close, within an LSOA that scores 42 (out of 100). Given the closure will only reduce the AvCash score of an infrastructure rich LSOA by 1, it's unlikely to affect localised financial exclusion. However, TSB customers across the city will no longer have an easily accessible city-centre branch to access financial advice, whilst a branch may score low in an assessment of cash access, it can be a positive contributor to financial inclusion.

Though not supported by any empirical data in this study, the conversion of free ATM's to charging, as discussed in Tischer et al. (2020), could be concerning. Tischer et al. (2019) found 16 ATM's (largely owned by IADs) converted to charging in 13 months, disproportionately affecting deprived communities. Despite the concerning shift being attributed to LINK's reduction in interchange fee, COVID-induced footfall reductions could reduce the profitability of lonely free ATM's (as seen in Map 10). Ultimately, decisions on infrastructure withdrawal are subjective and should be taken alongside considering the needs of affected communities.

## Refining the Priority Areas and Developing Local Solutions

This research has shown the need to identify pockets of financial exclusion and the underlying causes, whether they be infrastructure or (geo)demographic based. Though this study questioned the inclusion of Clifton and exclusion of Bulwell as priority areas, it was agreed that Aspley and St. Ann's face a heightened financial exclusion risk. The Nottingham Financial Resilience Partnership is still developing policy plans (Nottingham Financial Resilience Partnership, 2020), based on local needs and provisions. Shared priorities include improving access to credit unions, tackling the ID barrier in opening bank accounts, and expanding financial education.

Purely taking an infrastructure perspective would ignore the differing needs across the population, whilst others may have more pressing issues such as low incomes making savings and insurance unaffordable.

## Understanding the Effect of Post-Covid-19 Socio-Economic Changes

Relationships with money and financial infrastructure have been uncertain since March 2020, and the long-term impact of that is still unclear. Cummins (2020) found increased concern over financial wellbeing, with many struggling with reduced working hours or redundancies, and the anticipated effects of furlough being withdrawn are still unclear. Therefore, census data could be unreliable considering wider societal changes. Therefore, caution should be applied in trying to interpret long-term implications from data collected at a single point in time.

Given that city centre footfall has fallen due to COVID-19 (Centre for Cities, 2021), many free ATM's may now be unprofitable, and the loss or conversion of these machines could impact the index score. What is clear is that many ATM's are heavily reliant on passing trade, notablyin district centres, transport hubs and along major arterial routes. Despite the noted limitation of this study (that the framework doesn't consider the fact that many access cash on journeys to work or leisure), a shift of attention to residential based cash access could be a reasonable assumption to make in the light of increased (and flexible) home working. However, trends are showing a reduction in cash usage and an increase in online and card-based payment. Despite the partial re-opening in August 2021, cash use remained below half of the normal levels, whilst online alternatives for goods bought in physical stores remained a quarter higher (Davenport et al., 2020).

## Suggested recommendations and interventions

Thus far, the industry's response to infrastructure withdrawal has been to incentivise providers (particularly independent companies) to 'protect' their machines in areas where there is no nearby alternative (LINK, 2019). When a customer uses the machine, their bank pays a higher fee to the ATM provider as an incentive, however, with further reductions in ATM numbers, more protected ATM's are likely to increase the costs of the overall system. There also comes the risk of providers trying to game the system, by maximising the number of protected ATM's within their estate, thereby maximising the fees paid to them (Evans et al., 2020). Introducing recyclers could protect vulnerable infrastructure by improving its profitability. Despite there being many eligible ATM's to be protected, there are none in Nottingham covered by the scheme, they are either just over the city border or in rural villages (LINK, 2021). Therefore, three alternative interventions have been designed: [1] Offering Post Office style financial services at PayPoint's, [2] Introducing cash recyclers in place of lonely ATM's and [3] Improving digital inclusion.

| INTERVENTION 1: OFFERING FINANCIAL SERVICES AT PAYPOINTS | |
|---|---|
| Description | Offering Post Office style banking services at PayPoint's 27,000 outlets, currently being trialled in 13 stores elsewhere in the country (Lunn, 2020). |
| Areas benefitting | The effect appears greatest in areas of low financial inclusion (Arboretum, Bulwell Forest, St. Ann's), though the effect is weaker in some areas of Bestwood and Dales wards. See Map A in Appendix 1 demonstrating the change in index score. |
| Opportunities | <ul><li>Good geographic coverage (nearest neighbour score of 0.7).</li><li>Supports PayPoint's existing business model, providing a valuable community resource (parcels, bill payments).</li><li>Access to small balances in areas where an ATM may be unviable.</li></ul> |
| Challenges | <ul><li>Already low take-up of Post Office services (Parrott, 2018).</li><li>Post Offices have statutory obligations (Booth, 2019) whereas PayPoint's are franchises.</li><li>Significant fee for banks for a relatively unprofitable service (Thiel, 2008), they may prefer to shift customers to online banking with lower transaction costs (Peachey, 2019).</li><li>Doesn't address the issue of increasing participation in the financial system.</li></ul> |

*Table 8: Summary of Intervention 1*

The financial exclusion index presented earlier in this research was amended to take account of the intervention (Table 8), with Map A in the appendix showing the change in index rankings for LSOA's. PayPoint financial services increase scores by up to 2 points in St. Ann's, the Arboretum and Bulwell Forest. It appears to have the greatest effect in areas of low financial inclusion, though the effect is less strong in some areas of Bestwood and Dales ward. The fact that PayPoint's appear to be concentrated in areas of deprivation (Map 9), with existing strong take-up of bill payments and the benefits exceptions service could improve their chance of success.

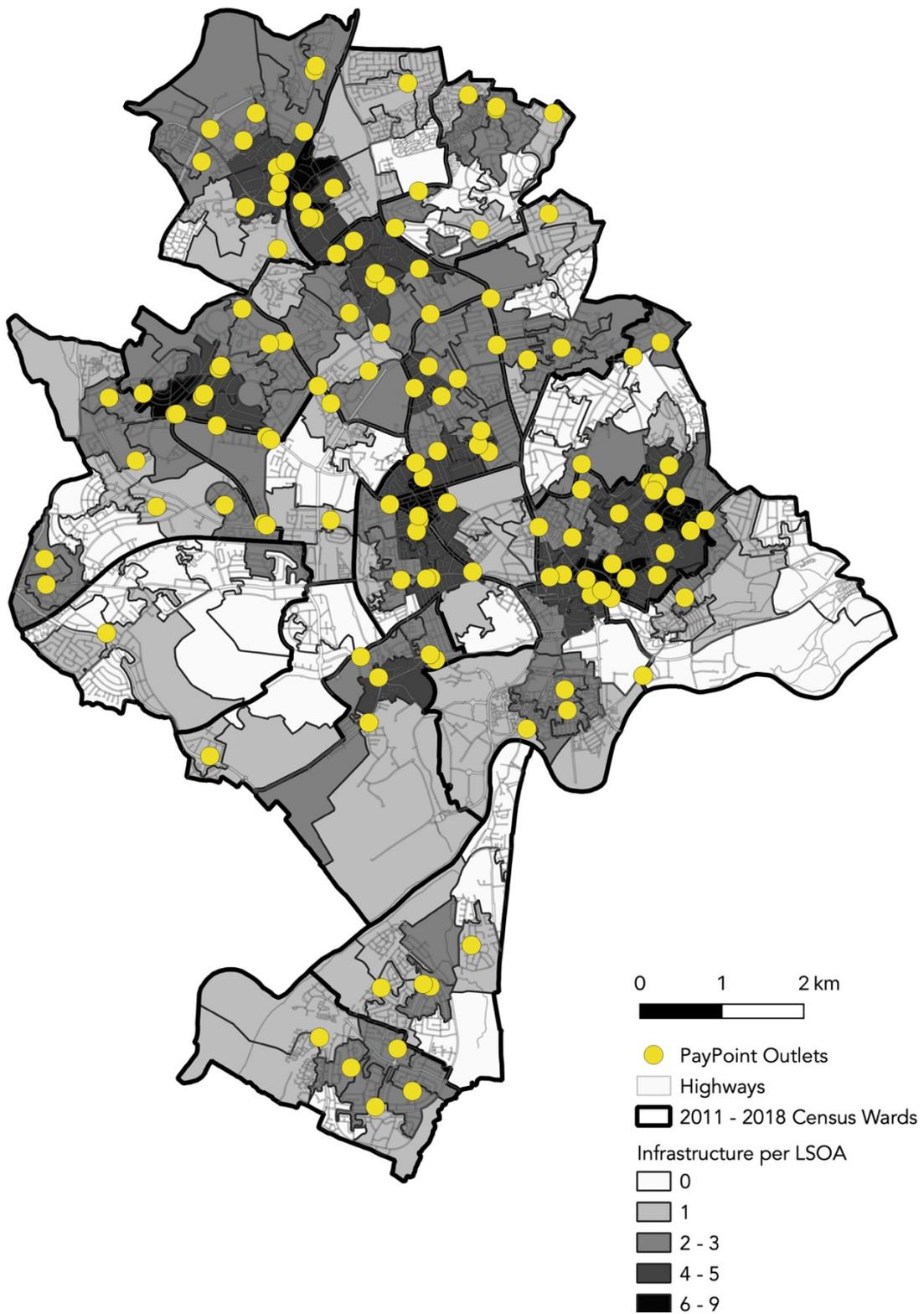

*Map 9: Distribution of PayPoint outlets*

INTERVENTION 2: CASH RECYCLERS

| | |
|---|---|
| Description | Lonely free ATMs could be converted to cash recyclers, whereby cash paid in is also paid back out (as in a traditional bank branch) (Automatia, 2021). |
| Areas benefitting | No apparent link with financial exclusion, with some of the loneliest ATMs located in areas of high financial inclusion. There are some benefits to Berridge, Arboretum, and Basford wards. See Map B in Appendix. |
| Opportunities | <ul><li>Reduces the likelihood of ATMs running out of cash.</li><li>More profitable, with a reduced need for cash in transit deliveries.</li><li>Greater range of services available, allowing consumers to better manage their finances.</li></ul> |
| Challenges | <ul><li>Recyclers are more expensive and represent a significant investment for operators if placed in an area of light use.</li><li>The effectiveness of the technology in encouraging greater ATM use is unproven, and there is likely to be safety concerns from consumers depositing large amounts of cash at an unmanned, often isolated machine.</li><li>Requires wider industry reform and the purchase of machines by banks from IADs.</li></ul> |

*Table 9: Summary of Intervention 2*

Converting lonely ATM's to recyclers (Table 9) brings benefits to Berridge, Arboretum, and Basford (Map B in appendix), as well as areas of high financial inclusion (Wollaton and Lenton). Whilst the benefits to an LSOAs AvCash score (and subsequently the financial exclusion score) are more pronounced than intervention 1, (a recycler has the same score as two Post Offices), Map 10 shows they are often located a greater distance from the population-weighted centroids than PayPoint's, so likely to benefit less of a population than intervention 1.

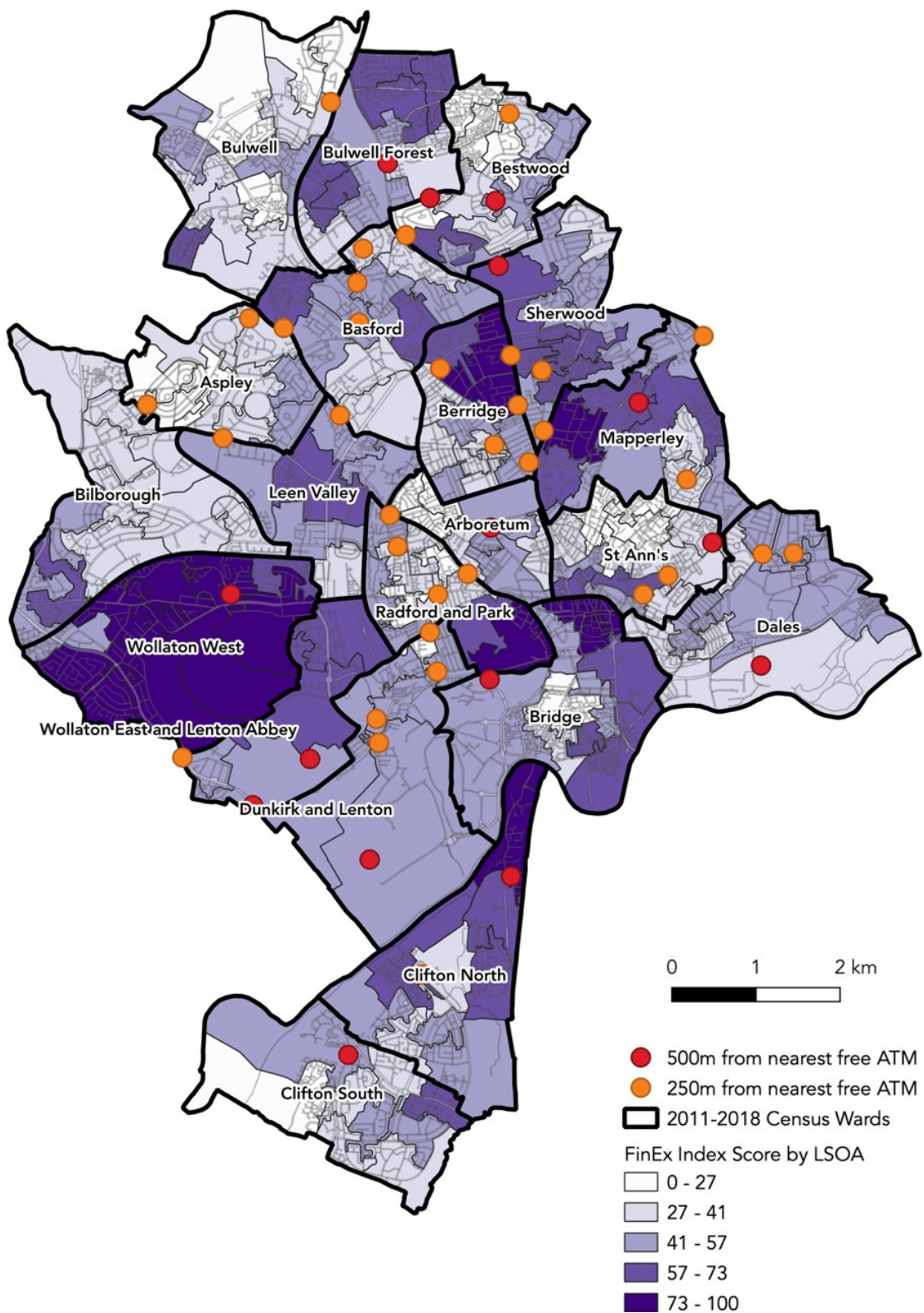

*Map 10: Potential ATMs to be converted to cash recyclers.*

INTERVENTION 3: IMPROVING DIGITAL INCLUSION

| | |
|---|---|
| Description | Uprating the LSOAs with an Internet User Classification score of 10 (e-withdrawn) to 7 (Passive and Uncommitted users), representing an improvement in digital engagement. |
| Areas benefitting | Largely in the north of the city (Aspley, Bilborough, Bulwell), with some pockets in St. Ann's and Clifton. There is little benefit to inner-city areas suffering financial exclusion (Radford, Lenton), as the young and professional populations are generally internet savvy. |
| Opportunities | <ul><li>Technology is becoming crucial for social interaction, access to education and managing money (Baker et al., 2020).</li><li>Opportunity to leverage the UK's innovative fintech sector to develop more inclusive payment systems.</li><li>Internet access could be provided as an essential utility, like electricity, with a trusted location in every local area (possibly a Post Office) to support inclusive internet access (Good Things Foundation, 2020).</li><li>Could be funded through the dormant assets scheme (DCMS et al., 2018) and delivered by the local charitable sector, in a similar way to Cambridge's New Horizons programme (Burgess, 2020).</li><li>Can address wider societal issues, such as money management and unemployment.</li></ul> |
| Challenges | <ul><li>High capital cost, but could be reduced through using laptops loaned to families on low incomes during the second national lockdown.</li><li>Likely to be a significant, expensive and long-term undertaking, even when targeting just the e-withdrawn (10% of LSOAs).</li><li>Strong existing fear of technology and distrust of banks amongst some populations (Britain Thinks, 2019).</li></ul> |

*Table 10: Summary of Intervention 3*

The most comprehensive effect on financial inclusion is improving digital adoption (Table 10). Uprating an LSOAs' IUC score to e-mainstream can result in as great as a 34-point increase in financial inclusion (see Map C in appendix). The e-withdrawn have the lowest rate of engagementin information searching and financial services online and rarely use the internet on a mobile (Alexiou and Singleton, 2018). Though broadband access may not be the issue, as there is an above-average take-up of cable TV, reducing the investment needed to bring these users online. The benefit is felt strongest in the north of the city (Aspley, Bilborough, Bulwell), with some pockets in St. Ann's and Clifton (Map 11).

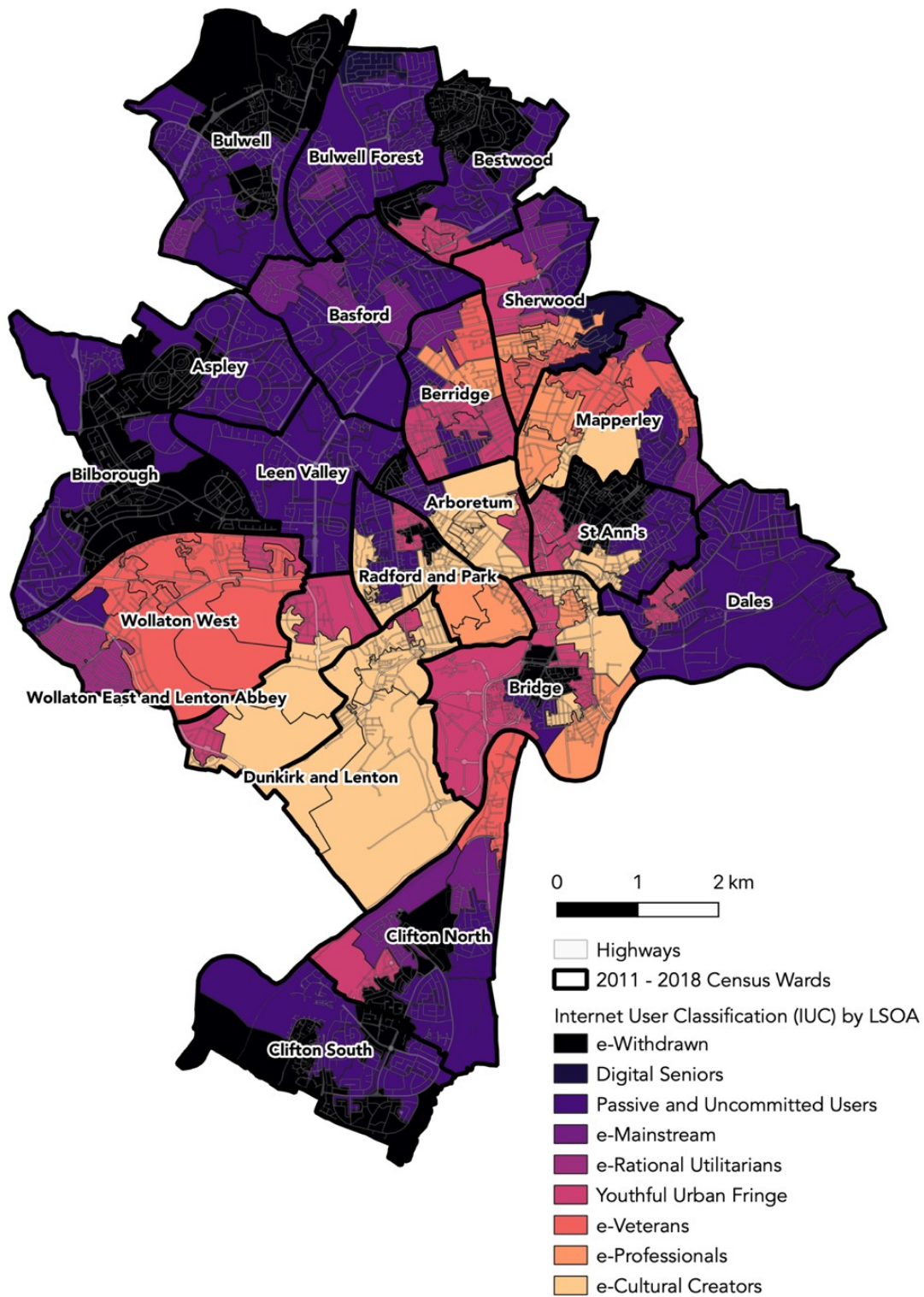

*Map 11: Internet User Classification by LSOA.*

The greatest intervention to improve financial inclusion is likely to be improving digital adoption, though it is likely to be a substantial, expensive, and long-term undertaking. Introducing recyclers bring some benefits, though they benefit relatively small populations, and targeting the loneliest ATM's doesn't tackle financial exclusion directly (the impact on the index score of incorporating *all* interventions can be seen in Map D in the appendix). This strategy could be effective in deprived rural areas that have seen bank branches close. As a short-term intervention, offering banking at PayPoint's appears the most effective. The infrastructure is largely in place, outlets are prevalent in areas of financial exclusion and consumers are accustomed to completing some financial transactions there. Supported by effective implementation and marketing, this appears to deliver a rapid and effective step towards financial inclusion.

This research and that of Sliced Bread Consulting et al. (2015) has demonstrated that LINK's model of protecting and installing ATM's on a distance measure will not be adequate to reduce financial exclusion. Vulnerable communities often lack the knowledge or resources to apply for one, whilst isolated ATM's come with further risks. Currently, there is little regulatory pressure for banks to actively address financial exclusion, and communities in the most need are often untouched by private businesses, as the needs are unprofitable (Duncan and Mary, 2008). There has also been little progress in the last five years to reduce the number of unbanked (House of Lords Liason Committee, 2021). Whilst the outlook seems to be the Financial Conduct Authority taking greater responsibility for regulating a commitment to financial inclusion (Ceeney, 2019), progress appears to be slow. In the meantime, Nottingham is in a unique position with its existing commitment to partnership working to deliver change through the Financial Resilience Partnership. No organisation alone has the power or responsibility to reduce financial exclusion, the financial system can maintain the current level of infrastructure provision, with local organisations helping smooth the transition to a digital world, tackling the barriers to participation. However, regulatory impetus is required on both sides to allow this to happen. Financial exclusion should be seen as a market failure, and firms alone cannot address the inequalities, but wider government intervention is needed.

# Research Summary

This applied research aimed to develop an area-based index of financial exclusion, considering infrastructure provision, (geo)demographics and the availability of alternatives, considering potential infrastructure withdrawal as Britain becomes cashless. It began with a comprehensive review of the literature, understanding that the indicators of financial exclusion are contested, and cash infrastructure is undergoing rapid changes as British society shifts to becoming cashless. A comprehensive, replicable, three-dimensional financial exclusion index was constructed using freely available data, the key finding being that the geographic distribution of the financially excluded population is not simple. It is complex and nuanced, in some area's infrastructure will play a bigger role, in others, (geo)demographics and poverty are likely to be a greater influence. Similarly scoring areas have distinctly different characteristics and challenges, purely protecting cash access based on a distance measure is does not appear to be an effective policy position. There is a strong link between social and financial exclusion, areas classed as hard-pressed communities and multicultural living appear to have lower financial inclusion scores. Policies solely targeting infrastructure provision or purely addressing social exclusion are unlikely to be effective. An integrated, partnership model is needed, drawing on the expertise of local communities and civil society. National regulation coupled with local intervention, where the provision of financial services is considered as a necessary utility, is needed.

Ultimately, index composition and weightings are subjective, it may be the case in certain areas that other influences are more pronounced (e.g. age, health, educational attainment), or consumers have different preferences for accessing cash. The measurement of infrastructure from residential areas does not consider the flow of residents between areas for work or leisure, and whilst mobility is measured somewhat by including car availability, other factors need to be considered. A network analysis could have been used to produce more accurate results; however, this would add additional complexity to an already comprehensive index (McEntee and Agyeman, 2010), and this was avoided given the reproducibility remit. Though the distance of 500m is supported by Clarke et al. (2002), it is subjective, and Evans et al. (2020) have adopted a 1km measure for rural areas. Additionally, Nottingham's city boundary does not cover the whole urban area, so the index does not count heavily used and accessible infrastructure just over the border, skewing results for LSOAs at the boundary. Whilst the majority of the data came from the census due to its reliability and replicability, the data is now approaching 10 years old, and though more recent data sources could be used they are often not at the resolution needed for this analysis. However, the imminent release of 2021 census data will enable the index to be updated.

The key learning from the research is that changes are needed in how financial infrastructure is provided, to better consider financial inclusion. A community-based perspective to understanding infrastructure provision is required, with the involvement of councils, charities and residents. As such, this index should be used by local authorities (with adjusted weightings to better reflect their locality) and banks to evaluate the impact of infrastructure withdrawal or identify locations for community banking outreach (Megaw, 2020). However, further statistical analysis is needed to better understand the causes of financial exclusion in 2021. Whilst offering financial services at PayPoint outlets can be an effective short-term intervention, a long-term comprehensive regulatory re-think is needed to maintain current levels of provision and smooth the transition towards a digital future. This should also consider the outcome of the community access to cash pilots (Lunn, 2020).

Further research is needed to develop a sustainable financial funding model that tackles financial exclusion and ensures an appropriate level of financial infrastructure is provided in the right places. However, the constraints of a commercially driven financial sector means a partnership approach is necessary. Research could also be undertaken to understand the level of financial exclusion across the UK, whilst a longitudinal study looking at changes in infrastructure provision would be particularly useful given the anticipated changes post- COVID-19.

# Appendix

Supplementary maps detailing changes in financial exclusion index score after interventions Implemented:

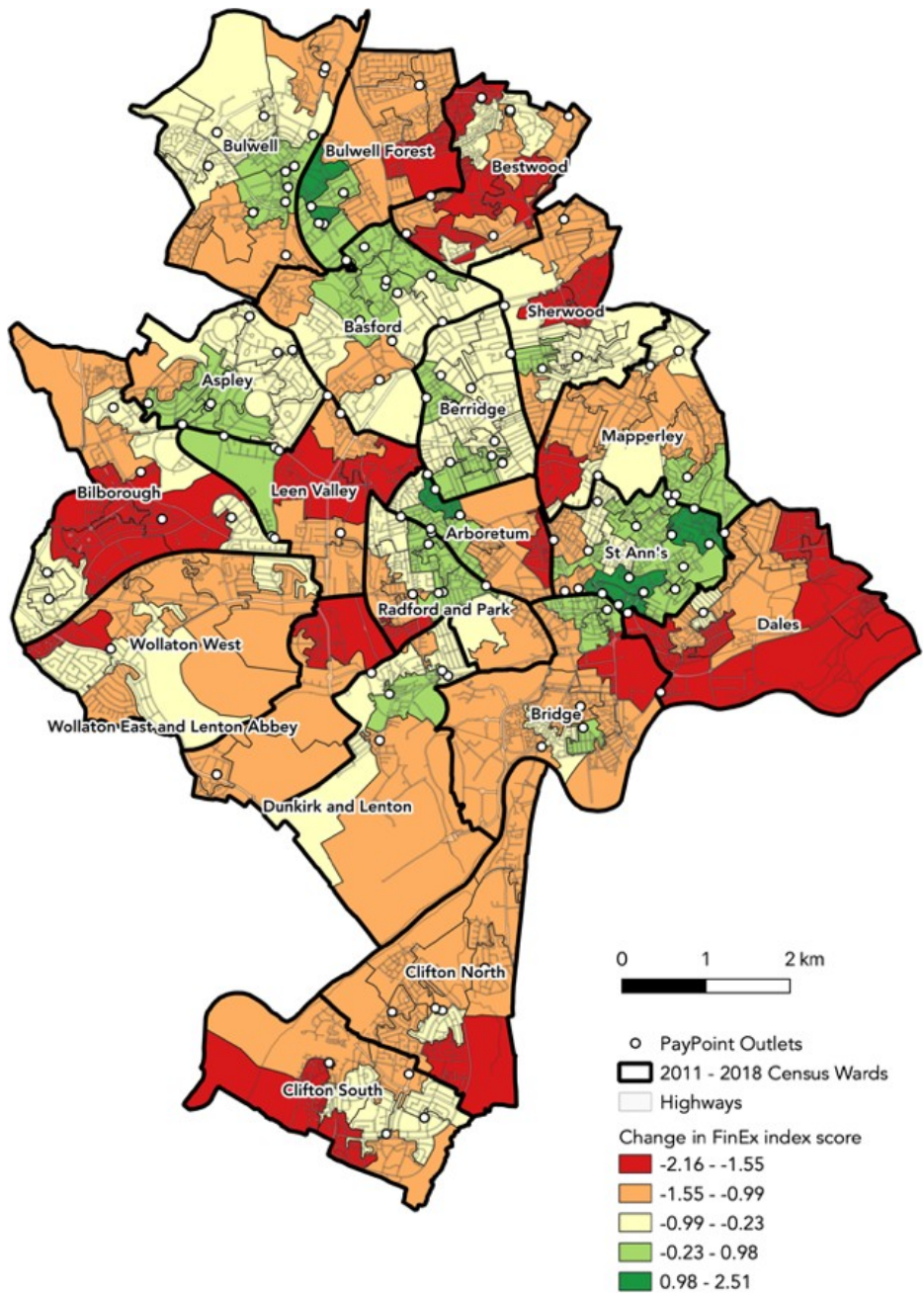

*Map A: Change in FinEx Index score after implementing Intervention 1*

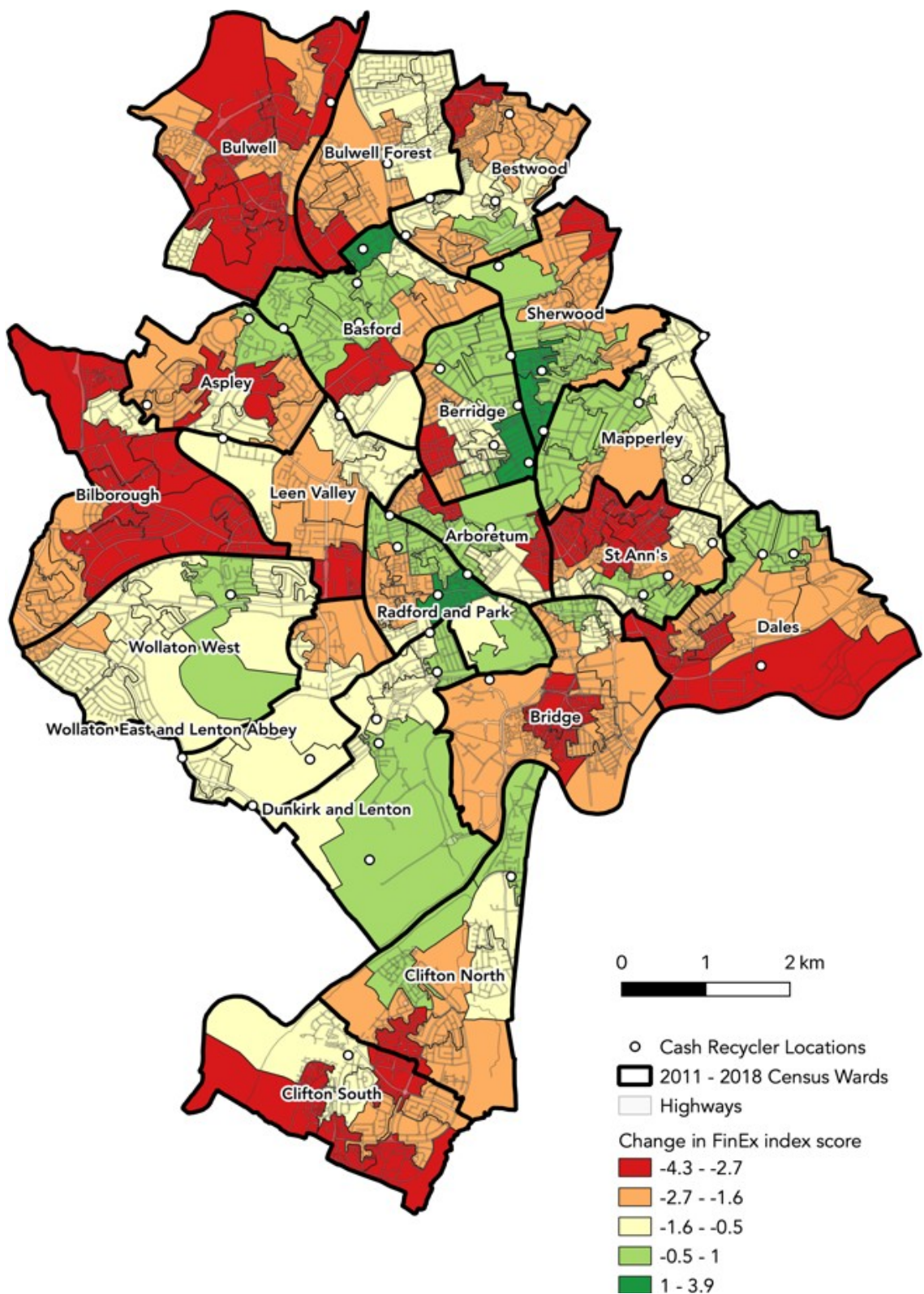

*Map B: Change in FinEx Index score after implementing Intervention 2*

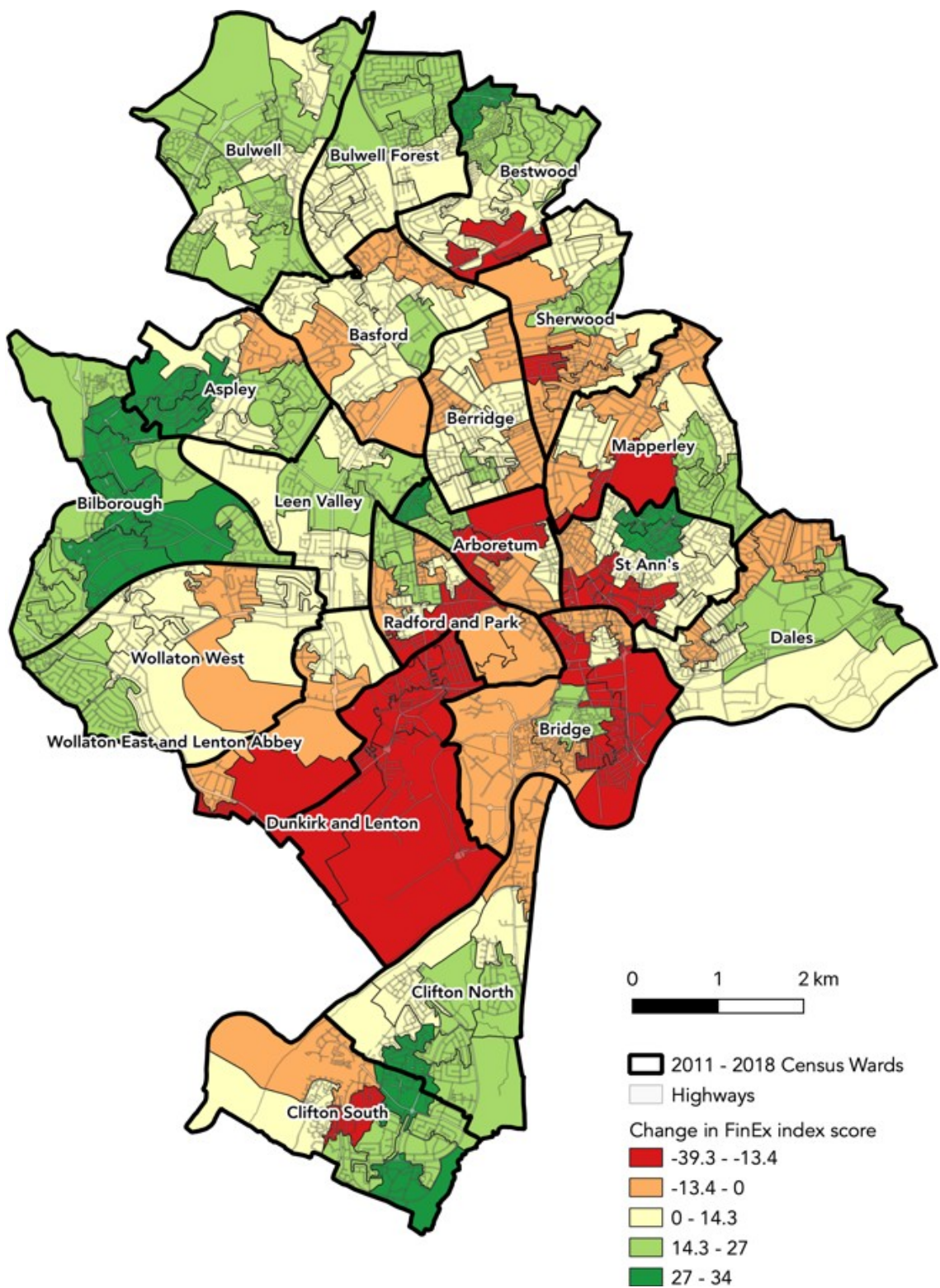

*Map C: Change in FinEx Index score after implementing Intervention 3*

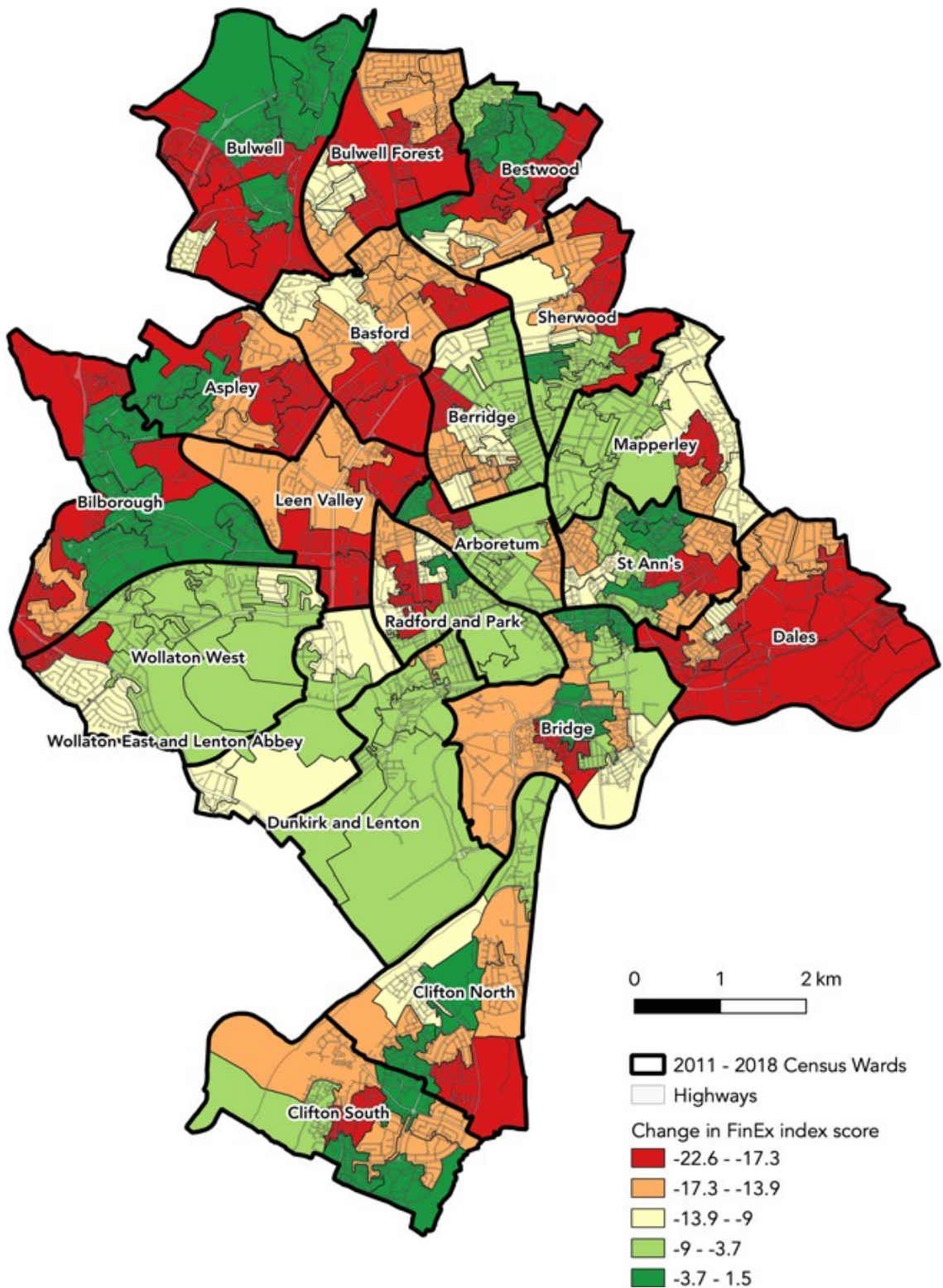

*Map D: Change in FinEx Index score after implementing all interventions*

**George Sullivan & Luke Burns**
University of Leeds, Leeds, United Kingdom
Corresponding author: Dr Luke Burns (l.p.burns@leeds.ac.uk).